\documentclass[fleqn,usenatbib]{mnras}
\usepackage{newtxtext,newtxmath}
\usepackage[T1]{fontenc}
\DeclareRobustCommand{\VAN}[3]{#2}
\let\VANthebibliography\thebibliography
\def\thebibliography{\DeclareRobustCommand{\VAN}[3]{##3}\VANthebibliography}
\usepackage{graphicx}	
\usepackage{amsmath}
\newcommand{\nv}{\hat{\mathbf{n}}}

\title[Probing baryonic feedback with FRBs]{Probing baryonic feedback with fast radio bursts: joint analyses with cosmic shear and galaxy clustering}

\author[A. Wayland et al.]{
Amy Wayland,$^{1}$\thanks{E-mail: amy.wayland@physics.ox.ac.uk}
David Alonso,$^{1}$
and Robert Reischke$^{2}$
\\
$^{1}$Department of Physics, University of Oxford, Denys Wilkinson Building, Keble Road, Oxford OX1 3RH, United Kingdom\\
$^{2}$Argelander-Institut für Astronomie, Universität Bonn, Auf dem Hügel 71, D-53121 Bonn, Germany
}

\date{Accepted XXX. Received YYY; in original form ZZZ}
\pubyear{\the\year{}}

\begin{document}
\label{firstpage}
\pagerange{\pageref{firstpage}--\pageref{lastpage}}
\maketitle

\begin{abstract}
  Cosmological inference from weak lensing (WL) surveys is increasingly limited by uncertainties in baryonic physics, which suppress the non-linear matter power spectrum on small scales. Multi-probe analyses that incorporate complementary tracers of the gas distribution around haloes offer a pathway to calibrate these effects and recover unbiased cosmological information. In this work, we forecast the constraining power of a joint analysis combining fiducial data from a Stage-IV WL survey with measurements of the dispersion measure from fast radio bursts (FRBs). We evaluate the ability of this approach to simultaneously constrain cosmological parameters and the astrophysical processes governing baryonic feedback, and we quantify the impact of key FRB systematics, including redshift uncertainties and source clustering. We find that, even after accounting for these effects, a 3$\times$2-point analysis of WL and FRBs significantly improves cosmological constraints, reducing the degradation factor on $S_8$ by $\sim 80\%$ compared to WL alone. We further show that FRBs alone are sensitive only to a degenerate combination of the key baryonic parameters, $\log_{10} M_{\rm c}$ and $\eta_{\rm b}$, and that the inclusion of WL measurements breaks this degeneracy. Finally, we extend our framework to incorporate galaxy clustering measurements using Luminous Red Galaxy and Emission Line Galaxy samples, performing a unified 6$\times$2-point analysis of WL, dispersion measures of FRBs, and galaxy clustering. While this combined approach tightens constraints on $\Omega_{\rm m}$ and $\log_{10} M_{\rm c}$, it does not lead to a significant improvement in $S_8$ constraints beyond those obtained from WL and FRBs alone.
\end{abstract}

\begin{keywords}
cosmology: large-scale structure of Universe.
\end{keywords}



\section{Introduction}
 Weak gravitational lensing (WL) provides a powerful and unbiased probe of the total matter distribution in the Universe. By statistically measuring the coherent distortions of background galaxy shapes, WL traces the integrated mass along the line-of-sight, providing a direct measurement of the underlying matter field. As current and upcoming surveys, such as Euclid \citep{Euclid2011}, and the Vera C. Rubin Observatory \citep[LSST][]{Mandelbaum2018lsst, LSST2019}, achieve percent-level statistical precision, their cosmological constraining power will increasingly depend on how accurately theoretical models capture the non-linear growth of structure. On small scales, however, the matter distribution is significantly affected by baryonic physics, which suppresses the matter power spectrum by up to $30\%$ \citep{vanDaalen2011effects, Chisari2019modelling}. These processes, including star formation and feedback from active galactic nuclei (AGN), redistribute the gas within and around haloes. If unaccounted for, baryonic effects can introduce significant biases in cosmological parameter inference from WL data \citep[e.g.][]{Semboloni2011quantifying, Chisari2019modelling, Arico2023des, Garcia2024cosmic, Bigwood2024weak}.

 One strategy to mitigate the uncertainties arising from baryonic feedback is to apply conservative scale cuts, removing the small-scale data most affected by these processes \citep[e.g.][]{Prat2021dark, Zacharegkas2022dark, Amon2023consistent, Lange2023constraints}. However, this approach inevitably reduces the statistical power of WL measurements, highlighting the need for methods that can retain small-scale information. Hence, accurate models of baryonic physics are essential to fully benefit from the information contained in small-scale measurements. For example, hydrodynamical simulations provide a direct means of studying the complex interactions between baryons and dark matter, however, they rely on assumptions about sub-grid astrophysical processes \citep[e.g.][]{Dubois2014dancing, Somerville2015physical, McCarthy2016bahamas, Schaye2023flamingo}. In contrast, recent WL analyses including the Kilo-Degree Survey (KiDS) and Hyper Suprime-Cam (HSC) address uncertainties surrounding baryonic feedback by modelling small-scale effects with physics-based models, such as \texttt{HMcode} \citep{Mead2021hmcode, Dalal2023hsc, Abbott2023des}. Alternatively, observational tracers of the LSS can be used to constrain baryonic feedback and break degeneracies between astrophysical and cosmological parameters. Examples include the kinetic (kSZ) and thermal (tSZ) Sunyaev-Zel'dovich effect \citep{Sunyaev1972observations, Sunyaev1980velocity}, as demonstrated in studies such as \citet{Schaan2021act, Amodeo2021act, Troster2022joint, Bigwood2024weak, Kovac2025baryonification,2512.02954,Wayland2025calibrating}, as well as X-ray data \citep[e.g.][]{1911.08494,2309.02920,Ferreira2023xray,LaPosta2025insights}. Recently, fast radio bursts (FRBs) have emerged as a promising new tracer of baryonic structure, offering a complementary perspective on feedback processes and the distribution of diffuse gas.

 The dispersion measure (DM) extracted from FRBs traces the integrated column density of free electrons along the line-of-sight, and therefore maps the large-scale distribution of free electrons in the intergalactic medium (IGM). In contrast to the SZ and X-ray signals, which probe dense virialised regions, the DM of FRBs is most sensitive to the full complement of diffuse gas that dominates the cosmic baryon budget. Moreover, since the DM depends only on the electron number density and not on the gas temperature, FRBs probe baryons in a different way from other observables (e.g. tSZ or X-rays), which depend on both density and temperature. Previous studies using FRB samples have demonstrated the potential of this technique for cosmology and baryonic physics \citep[e.g.][]{1901.02418, Reischke2020primordial, 2102.11554, 2104.04538, Alonso2021linear, 
 2404.06049, 2403.08611, Reischke2024analytical}. For instance, \citet{Macquart2020census} used localised FRBs to confirm that the majority of the previously ``missing'' baryons reside in the diffuse IGM, consistent with predictions from hydrodynamical simulations \citep[e.g.][]{Sorini2022baryons, Veenema2026modelling}. Cross-correlation approaches have also been proposed that link the DM of FRBs with galaxy surveys and LSS tracers to test models of baryonic feedback \citep{2201.04142, Reischke2023calibrating, Sharma2025probing, Reischke2025first}. Moreover, simulations such as CAMELS have been employed to explore how different baryonic feedback prescriptions impact FRB statistics and their inferred dispersion measures \citep[e.g.][]{Medlock2024probing, Medlock2025constraining}. As the number of localised FRBs continues to increase rapidly with instruments such as CHIME/FRB \citep{Wang2025measurement, Abbott2026second} and DSA-2000 \citep{Hallinan2019dsa}, DM maps derived from these surveys will provide a new insight into baryonic physics on cosmological scales. Indeed, the first tentative detections of a correlation between FRB DM measurements and tracers of the large-scale structure have recently been made \citep{Wang2025measurement,2511.02155}.

 In previous work \citep{Wayland2025calibrating}, we determined that a combination of kSZ measurements and X-ray cluster gas fractions would allow LSST-like experiments to self-calibrate baryonic effects almost perfectly. These observables probe the thermal and dynamical states of the ionised gas within haloes, providing complementary information on the small-scale baryonic physics that drives feedback. However, including additional tracers of the large-scale structure (LSS) with complementary sensitivity to baryonic physics could allow us to break residual parameter degeneracies, and would constitute a vital alternative route to self-calibration, confirming the soundness and consistency of the physical model used to describe feedback. 
 
 In this work, we consider the DM of FRBs as such a tracer. Specifically, we extend the methodology of \citet{Wayland2025calibrating} by incorporating FRBs as an external tracer of baryonic structure, and perform a joint analysis of synthetic LSST-like weak lensing and FRB DM data in a 3$\times$2-point framework. The resulting auto- and cross-correlations between these observables provide a powerful means of connecting the total matter field traced by WL to the baryonic component traced by FRBs. The work presented here goes beyond the analysis of earlier studies \citep[e.g.][]{Reischke2023calibrating, Sharma2025probing} in several important directions. Firstly, we incorporate several additional FRB-related systematics that were not explored in the previous work. These include uncertainties in the FRB redshift distribution, the impact of source clustering between FRBs and their host galaxies, and the application of scale cuts to mitigate contamination from the Milky Way foreground. Moreover, we build upon the work of \citet{Reischke2020primordial}, which investigated how tomographic analyses of the angular DM correlation function can be used to constrain the primordial bispectrum shape parameter, $f_{\rm NL}$. In this work, however, we investigate FRB tomography in the context of baryonic feedback. Specifically, we assess how redshift binning of the FRB sample affects the ability to constrain a potential redshift dependence of the key baryonic parameter $\log_{10} M_{\rm c}$. We also explore the improvement in constraining power achieved when using a more futuristic FRB sample with an order-of-magnitude increase in the FRB number density. Furthermore, we directly compare the constraining power of FRBs on key baryonic parameters with that of kSZ and X-ray measurements, highlighting the complementary roles of these probes in constraining baryonic feedback.

 Additionally, galaxy clustering (GC) can be used as a local tracer of the underlying matter density field, offering a promising avenue for enhancing cosmological constraints by breaking parameter degeneracies. We therefore extend our analysis by incorporating galaxy clustering into a unified 6$\times$2-point framework together with WL and FRB DMs. While the combination of galaxy clustering and FRBs has previously been explored in the context of baryonic effects \citep{Sharma2025probing}, and has recently led to the first detection of a correlation between FRB DMs and the LSS \citep{Wang2025measurement}, our approach advances this framework by considering multiple galaxy populations with distinct halo occupation distributions, specifically, Luminous Red Galaxies (LRGs) and Emission Line Galaxies (ELGs). The inclusion of two types of galaxy samples is novel compared to existing studies involving FRBs and galaxy clustering \citep[e.g.][]{Sharma2025probing}, and allows us to directly probe the halo mass dependence of the bound gas fraction. In turn, this enables more stringent constraints to be placed on the baryonic processes governing the gas within haloes and helps to break degeneracies between astrophysical and cosmological parameters. The joint analysis of WL, FRBs DMs, and GC therefore provides a comprehensive, data-driven framework for calibrating baryonic feedback models, ultimately improving the robustness of cosmological constraints derived from WL measurements, both alone and in combination with clustering.

 The structure of the paper is as follows. In Section \ref{sec:model}, we introduce the theoretical model used to describe cosmic shear, FRB DM, galaxy clustering, and baryonic effects. The results of our analysis are presented in Section \ref{sec:results}, where we forecast the level to which we can recover cosmological constraints using our multi-tracer approach and the effect of FRB systematics. Finally, we conclude with a summary of our key findings and their implications for future cosmological surveys and multi-tracer analyses in Section \ref{sec:conclusions}.

\section{Large-scale structure tracers} \label{sec:model}
 \subsection{Power spectra} \label{ssec:model.spectra}
  In this work, we describe all LSS observables as projected fields on the sky, obtained by integrating 3D quantities along the line-of-sight with appropriate radial kernels. For any projected field $A(\nv)$, the spherical harmonic coefficients $a_{\ell m}^A$ are defined via
  \begin{equation}
      A(\nv) = \sum_{\ell m} a_{\ell m}^A \, Y_{\ell m}(\nv).
  \end{equation}
  Assuming statistical isotropy, the spherical harmonic coefficients of two projected fields $A(\nv)$ and $B(\nv)$ satisfy
  \begin{equation}
      \langle a_{\ell m}^A a_{\ell' m'}^{B*} \rangle = \delta_{\ell\ell'}^{\mathcal{K}} \delta_{mm'}^{\mathcal{K}} C_\ell^{AB},
  \end{equation}
  where $C_{\ell}^{AB}$ is the angular cross-power spectrum and $\delta_{ij}^{\mathcal{K}}$ denotes the Kronecker delta function. The resulting angular power spectrum can be expressed in terms of the 3D power spectrum $P_{AB}$ of the underlying fields as
  \begin{align}
      C_\ell^{AB} = \frac{2}{\pi} &\int_0^\infty \mathrm{d}k \, k^2 \int_0^{\chi_{\rm H}} \mathrm{d}\chi \, W_A(\chi) j_{\ell}(k\chi) \nonumber \\
      \times&\int_0^{\chi_{\rm H}} \mathrm{d}\chi' \, W_B(\chi') j_{\ell}(k\chi') P_{\rm AB}(k, z(\chi), z(\chi')),
  \end{align}
  where $W_A(\chi)$ and $W_B(\chi)$ are the radial kernels associated with the two tracers, $j_\ell$ is the spherical Bessel function of order $\ell$, and $\chi_{\rm H}$ denotes the comoving horizon distance. Under the Limber approximation \citep{Limber1954analysis, LoVerde2008extended}, we may write the angular power spectrum as
  \begin{equation} \label{eq:cl_ab}
      C_\ell^{AB} = \int_0^{\chi_{\rm H}} \mathrm{d}\chi \, \frac{W_A(\chi) W_B(\chi)}{\chi^2} \, P_{AB}\left(k=\frac{\ell+1/2}{\chi}, \, z(\chi)\right).
  \end{equation}
  Eq. \eqref{eq:cl_ab} provides the general framework for all angular auto- and cross-power spectra considered in this work. Specifically, we apply this formalism to cosmic shear ($\gamma$), FRB DMs ($\mathcal{D}$), and galaxy clustering ($\rm g$), and analyse the six spectra
  \begin{equation} \nonumber
      \left\{C_{\ell}^{\mathcal{DD}}, \, C_{\ell}^{\mathcal{D}\gamma}, \, C_{\ell}^{\gamma\gamma}, \, C_{\ell}^{\mathcal{D} \rm g}, \, C_{\ell}^{\rm g \gamma}, \, C_{\ell}^{\rm gg}\right\},
  \end{equation}
  which follow directly from Eq. \eqref{eq:cl_ab} for an appropriate choice of radial kernels and 3D power spectra. We specify these ingredients for the particular tracers under consideration in the following subsections. Noise contributions for each tracer (i.e. intrinsic shape noise for WL, DM/host variance for FRBs, and shot noise for GC) enter as additive diagonal terms of the form $N_{\ell}^{\rm A} \delta_{\rm AB}$.

  To compute the 3D power spectra $P_{AB}$ that enter Eq. \eqref{eq:cl_ab}, we use the Core Cosmology Library \citep[CCL][]{Chisari2019core}, which employs the halo model. Under this framework, the power spectrum is written as the sum of one-halo and two-halo contributions:
  \begin{equation}
      P_{AB}(k,z) = P_{AB}^{\rm 1h}(k,z) + P_{AB}^{\rm 2h}(k,z),
  \end{equation}
  where 
  \begin{equation}
      P_{AB}^{\rm 1h}(k,z) = \int_0^\infty \mathrm{d}M \, n(M) \, u_A(M,k) \, u_B(M,k)
  \end{equation}
  and
  \begin{equation}
      P_{AB}^{\rm 2h}(k,z) = P_{\rm lin}(k) \prod_{n=A,B} \left[\int_0^\infty \mathrm{d}M \, n(M) b(M) u_n(M,k) \right].
  \end{equation}
  Here, all halo model quantities are implicitly dependent on redshift $z$, and $u_A(M,k)$ is the spherically-averaged Fourier transform of the halo profile. This is in turn given by
  \begin{equation}
      u_A(M,k) = \int_0^\infty \mathrm{d}r \, 4\pi r^2 \frac{\sin(kr)}{kr} A_{\mathrm{h}}(r|M),
  \end{equation}
  where $A_{\mathrm{h}}(M,r)$ is the radial profile for the field $A$ in a host halo of mass $M$.
  
  We then evaluate all halo model components using the following parametrisations. We define halo masses using an overdensity threshold of $\Delta = 200$ relative to the critical density, $\bar{\rho}_{\rm c}$. For the structural and statistical properties of haloes, we adopt the concentration-mass relation $c(M)$ from \citet{Duffy2008dark}, the halo mass function $n(M)$ from \citet{Tinker2008toward}, and the halo bias $b(M)$ from \citet{Tinker2010large}. 

  One limitation of the standard halo model is its reliance on the linear bias assumption. This approximation leads to an underestimation of the matter power spectrum in the transition regime between the one- and two-halo terms. To mitigate the resulting biases in parameter inference, \citet{Mead2020including} developed a formalism that includes higher order bias terms. However, implementing this approach is beyond the scope of this work, and we therefore adopt the standard halo model.

 \subsection{Cosmic shear} \label{ssec:model.shear}
  Weak gravitational lensing arises from the deflection of photons as they travel from a distant background source towards the observer, induced by the gravitational potential of the intervening large-scale structure \citep{Bartelmann1999weak,Mandelbaum2017weak}. This effect produces coherent distortions in the observed shapes of background galaxies, which is known as cosmic shear. As an unbiased tracer of the underlying matter density field, cosmic shear provides a direct probe of the large-scale distribution of matter in the Universe. 
 
  We model the cosmic shear signal within the general angular power spectrum framework introduced in Section \ref{ssec:model.spectra}. In this context, the $E$-mode cosmic shear power spectrum between tomographic redshift bins $i$ and $j$ is obtained by evaluating Eq. \eqref{eq:cl_ab} with the 3D matter power spectrum, $P_{\mathrm{mm}}(k,z)$, and the lensing efficiency kernel, $W_{\gamma_i}(\chi)$, which in turn depends on the redshift distribution of the source, $p_i(z)$, via
  \begin{equation} \label{eq:wl_kernel}
    W_{\gamma_i}(\chi) = \frac{3}{2} G_\ell H_0^2 \Omega_{\mathrm{m}} \frac{\chi}{a(\chi)} \int_{z(\chi)}^\infty \mathrm{d}z' \, p_i(z') \frac{\chi(z') - \chi}{\chi(z')},
  \end{equation}
  where $a = 1/(1+z)$ is the scale factor. The prefactor $G_\ell$ is a purely geometric factor that accounts for the conversion between the 3D Laplacian of the gravitational potential and the angular Hessian of the corresponding lensing potential, arising from the spin-2 nature of the cosmic shear field and ensuring the correct normalisation of the cosmic shear power spectrum in harmonic space \citep{Kilbinger2017precision}. This factor is given by
  \begin{equation}
     G_\ell \equiv \sqrt{\frac{(\ell+2)!}{(\ell-2)!}} \frac{1}{(\ell+1/2)^2}.
  \end{equation}

  Similarly to \citet{Wayland2025calibrating}, we follow the specifications of the LSST Dark Energy Science Collaboration Science Requirements Document \citep{Mandelbaum2018lsst} for the 10-year weak lensing survey to construct our mock cosmic shear data. We adopt the source number density, redshift distribution, and photometric redshift uncertainty described in \citet{Leonard2023n5k}, dividing the sample into five tomographic bins of equal effective number density. The resulting cosmic shear data vector then consists of all unique auto- and cross-power spectra between these redshift bin pairs. We account for shot noise in the WL auto-power spectra by including the contribution $N_{\ell}^{\gamma\gamma} = \sigma_{\epsilon}^2 / \bar{n}_{\gamma}$, where $\bar{n}_{\gamma}$ is the galaxy number density of the WL redshift bin under consideration and $\sigma_{\epsilon} = 0.28$ is the single component ellipticity dispersion.

  \subsubsection{Systematics affecting cosmic shear} \label{sssec:model.shear.systematics}
   The cosmic shear signal is subject to several sources of systematic uncertainty, which can be broadly divided into two classes: calibratable and non-calibratable systematics. Calibratable systematics can be constrained through independent external data or instrument calibration, allowing tight priors to be imposed. In contrast, non-calibratable systematics must be constrained directly by the cosmic shear data.

   An important non-calibratable systematic in cosmic shear analyses is the intrinsic alignment (IA) of galaxies with the surrounding LSS \citep{Brown2002measurement}. The non-linear linear alignment (NLA) model is the simplest physical description for this effect \citep{Hirata2004intrinsic}. In this framework, the IA contribution is parametrised by a redshift-dependent amplitude,
   \begin{equation} \label{eq:NLA_IA}
       A_{\mathrm{IA}}(z) = A_{\mathrm{IA},0} \left(\frac{1+z}{1+z_*}\right)^{\eta_{\mathrm{IA}}},
   \end{equation}
   where $A_{\mathrm{IA},0}$ and $\eta_{\mathrm{IA}}$ are free parameters describing the normalisation and slope of the redshift power-law, respectively. We adopt a pivot redshift of $z_* = 0.62$, following \citet{DES:2017troxel} and \citet{DES:2017abbott}.

   Cosmic shear measurements are also significantly impacted by photometric redshift uncertainties and multiplicative shape biases \citep{Bonnett2016redshift, Hildebrandt2020cosmic},  both of which are examples of calibratable systematics. Photometric redshift uncertainties arise from imperfect knowledge of the true redshift distributions in photometric redshift surveys. We model this effect by introducing a shift parameter $\Delta z_i$ for each tomographic bin, such that the true redshift distribution is given by $p_i(z) = \hat{p}_i(z+\Delta z_i)$, where $\hat{p}_i(z)$ is the best-guess redshift distribution \citep{Ruiz-Zapatero:2023}. 
   
   Multiplicative shape biases  arise due to limitations in image resolution and noise, which affect galaxy shape measurements \citep{Miller2013bayesian}. These biases are parametrised by a multiplicative factor $m_i$ for each redshift bin $i$ \citep[see e.g][]{Hildebrandt2017cosmological}. As a result, the observed angular power spectra, $\widetilde{C}_\ell$, are related to the true angular power spectra, $C_{\ell}$, via
   \begin{equation}
       \widetilde{C}_{\ell}^{ij} = (1+m_i)(1+m_j) \, C_{\ell}^{ij},
   \end{equation}
   where $m_i$ and $m_j$ are the multiplicative biases of the redshift bins $i$ and $j$, respectively.
   
   We marginalise over the calibratable systematics using the analytical approximation introduced by \citet{Hadzhiyska2023cosmology} and \citet{Ruiz-Zapatero:2023}, which is based on the Laplace approximation. This approach avoids explicitly introducing two additional nuisance parameters per redshift bin into the likelihood, leading to a substantial improvement in computational efficiency. In practise, analytical marginalisation simply amounts to updating the covariance matrix of the data, $\mathsf{C}$, to
   \begin{equation} \label{eq:updated_cov}
       \widetilde{\mathsf{C}} \equiv \mathsf{C} + \mathsf{T}\,\mathsf{P}\,\mathsf{T}^T,
   \end{equation}
   where $\mathsf{P}$ is the covariance matrix of the calibratable systematics, which we assume to be uncorrelated, and the matrix $\mathsf{T}$ contains the derivatives of the theoretical prediction $\mathbf{t}$ with respect to the set of calibratable parameters evaluated at their prior mean values. We assume Gaussian priors on both $\Delta z_i$ and $m_i$, with 68 per cent uncertainties on $\sigma(\Delta z_i) = 0.001 (1+\bar{z}_i)$ and $\sigma(m_i) = 0.01$, consistent with the LSST requirements reported in \citet{Mandelbaum2018lsst}.

 \subsection{FRB dispersion measure} \label{ssec:model.frb}
  \subsubsection{FRB statistics}
   As electromagnetic pulses propagate through an ionised medium composed of free electrons, their group velocity becomes frequency-dependent. This dispersion results in a frequency-dependent time delay that scales inversely with the square of the observed frequency \citep{2007.02886}. The proportionality constant, known as the dispersion measure (DM), quantifies the column density of free electrons along the line-of-sight.

   The total observed DM, $\mathrm{DM}_{\rm tot}$, can be expressed as the sum of three distinct contributions:
   \begin{equation}
       \mathrm{DM}_{\rm tot}(\nv,z) = \mathrm{DM}_{\rm LSS}(\nv,z) + \mathrm{DM}_{\rm MW}(\nv) + \frac{\mathrm{DM}_{\rm host}}{1+z},
   \end{equation}
   where $\mathrm{DM}_{\rm LSS}$ represents the contribution from the LSS of the Universe, $\mathrm{DM}_{\rm MW}$ denotes the Milky Way contribution, and $\mathrm{DM}_{\rm host}$ is the rest-frame DM associated with the host galaxy of the FRB.

   The LSS contribution, $\mathrm{DM}_{\rm LSS}$, encodes valuable cosmological information, as it traces the cosmic electron distribution, and can be written as
   \begin{equation} \label{eq:DM_LSS1}
       \mathrm{DM}_{\rm LSS}(\nv,z) = \int_0^z \mathrm{d}z' \, n_{\rm e,c}(\nv,z') \frac{1+z'}{H(z')},
   \end{equation}
   where $H(z)$ is the Hubble parameter and $n_{\rm e,c}$ is the comoving free electron number density. The physical electron number density is related to its comoving counterpart via $n_{\rm e} = n_{\rm e,c} a^{-3}$, and in turn depends on the local matter overdensity, $\delta_{\rm m}$, according to
   \begin{equation} \label{eq:n_e}
       n_{\rm e, c}(\nv,z) = \frac{(1+x_{\rm H}) \, \bar{\rho}_{\rm b}(z)}{2m_{\rm p}} \left[1+\delta_{\rm e}(\nv,z)\right],
   \end{equation}
   where $\Bar{\rho}_{\rm b}(z)$ is the mean baryon density, $m_{\rm p}$ is the proton mass, $x_{\rm H}=0.76$ is the hydrogen mass fraction, and $\delta_{\rm e}$ is the electron overdensity. The mean baryon density can be written as $\bar{\rho}_{\rm b} = \Omega_{\rm b0} \rho_{\rm crit,0}$, with $\rho_{\rm crit,0} = (3H_0^2)/(8 \pi G)$ denoting the present-day critical density. The cosmological contribution to ${\rm DM}_{\rm tot}$ is then
   \begin{equation} \label{eq:DM_LSS2}
     \mathrm{DM}_{\rm LSS}(\nv,z) = \mathcal{A} \int_0^z \mathrm{d}z' \, \frac{(1+z')}{H(z')} \left[1+\delta_{\rm e}(\nv,z')\right],
   \end{equation}
   where the prefactor ${\cal A}$ is 
   \begin{equation} \label{eq:prefactor_A}
     \mathcal{A} \equiv \frac{3 H_0^2 \Omega_{\rm b0}}{8 \pi G m_{\rm p}} \, \frac{(1 + x_{\rm H})}{2} \, (1-f_{\rm neutral}-f_{\rm star}).
   \end{equation}
   Here, $f_{\rm neutral}$ denotes the fraction of baryons in the form of neutral gas. Observations of the cosmic neutral hydrogen density indicate that $f_{\rm neutral} \sim 10^{-3}$ at low redshift \citep[e.g.][]{Jones2018alfalfa, Obuljen2019HI}. This contribution is therefore at the sub-percent level and has a negligible effect on our forecasts. Hence, for simplicity, we may drop this factor and assume fully ionised gas. We also subtract the stellar component, $f_{\rm star}$, to account for the fraction of baryons bound in stars, which do not contribute to the ionised gas. This fraction is given by the mass-function weighted average of Eq.~\eqref{eq:f_star}.

  \begin{figure*}
    \centering
    \includegraphics[width=\linewidth]{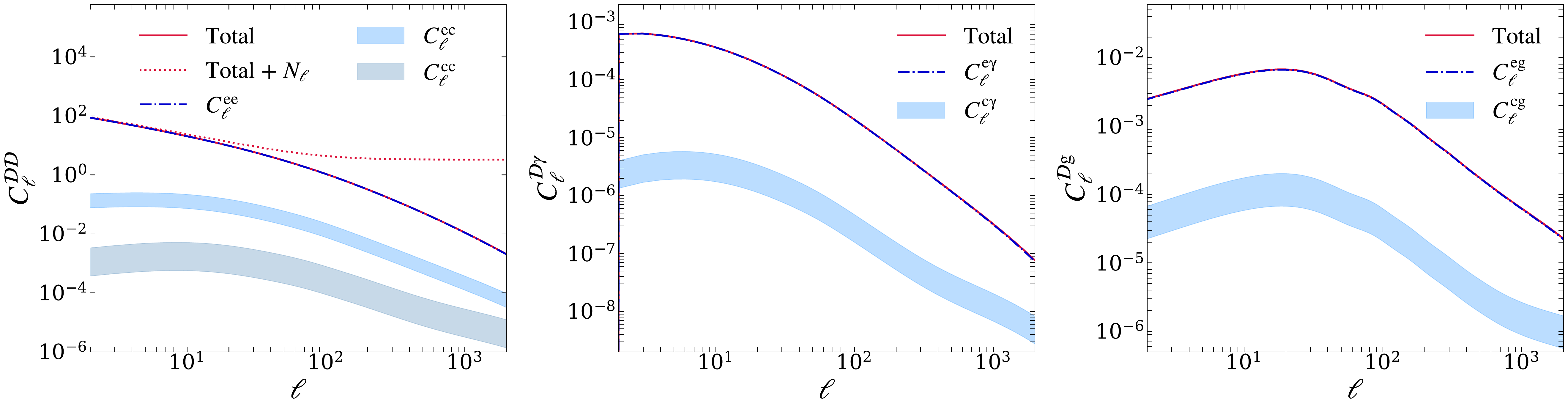}
    \caption{The contributions to the angular power spectrum arising from FRB source clustering effects. The bands corresponding to the correlations involving the source-clustering term represent the range $b_{\rm f} \in \{1.0, 3.0\}$ for the FRB bias. \textit{Left Panel.} Components of the FRB auto-correlation, $C_{\ell}^{\mathcal{DD}}$, as given in Eq. \eqref{eq:Cl_DD}. \textit{Middle Panel.} Components of the cross-correlation between the FRB DM and the first WL redshift bin, $C_{\ell}^{\mathcal{D}\gamma}$, from Eq. \eqref{eq:Cl_Dgamma}. \textit{Right Panel.} Components of the cross-correlation between the FRB DM and galaxy clustering for the LRG sample, $C_{\ell}^{\mathcal{D}\rm g}$, from Eq. \eqref{eq:Cl_Dg}.}
    \label{fig:cls_source_clustering}
   \end{figure*}
  
   Following \citet{Reischke2020primordial}, the LSS contribution to the total DM can be decomposed into a mean background term and the spatial fluctuations:
   \begin{equation}\label{eq:DM_mean_fluct}
     \mathrm{DM}_{\rm LSS}(\nv,z) = \overline{\mathrm{DM}}_{\rm LSS}(z) + \mathcal{D}(\nv,z),
   \end{equation}
   where $\mathcal{D}(\nv,z)$ denotes the perturbation to the DM induced by inhomogeneities in the free electron distribution (i.e. the contribution in Eq. \eqref{eq:DM_LSS2} proportional to $\delta_{\rm e}$). This fluctuation can be written as a weighted line-of-sight integral over the electron overdensity as follows. Given a normalised redshift distribution $n(z)$ of FRB sources, satisfying $\int \mathrm{d}z \, n(z) = 1$, and defining the corresponding distribution in comoving distance $\chi$ as
   \begin{equation}
       n(\chi) = n(z) \frac{\mathrm{d}z}{\mathrm{d}\chi},
   \end{equation}
   the redshift-averaged DM fluctuation is given by
   \begin{equation} \label{eq:effective_DM1}
       \mathcal{D}(\nv) = \int_0^{\chi_{\rm H}} \mathrm{d}\chi \, n(\chi) \mathcal{D}(\nv,z(\chi)),
   \end{equation}
   where $\chi_{\rm H}$ denotes the comoving horizon distance. Combining Eqs. \eqref{eq:DM_LSS2} and \eqref{eq:effective_DM1}, and after simple algebra, we obtain the following expression for the redshift-average DM fluctuation:
   \begin{equation} \label{eq:effective_DM2}
     \mathcal{D}(\nv) = \int_0^{\chi_{\rm H}} \mathrm{d}\chi \, W_{\mathcal{D}}(\chi) \delta_{\rm e}(\nv, z(\chi)),
   \end{equation}
   where the DM radial kernel is
   \begin{equation} \label{eq:frb_kernel}
     W_{\mathcal{D}}(\chi) = {\cal A}\,(1+z(\chi)) \int_{\chi}^{\chi_{\rm H}} \mathrm{d}\chi' n(\chi').
   \end{equation}

  \subsubsection{FRB source clustering} \label{sssec:model.frb.sc}
   As discussed in recent studies such as \citet{Alonso2021linear} and \citet{Wang2025measurement}, the observed DM correlations are not solely determined by fluctuations in the large-scale electron distribution. In particular, an additional contribution arises from the spatial clustering of FRB host galaxies, which modulates the observed DM field. Physically, this term captures the fact that FRBs preferentially occur in overdense regions that may also host an excess of free electrons. As a result, the observed DM fluctuations are partially sourced by the mean DM modulated by the FRB overdensity, and by correlations between the spatial distribution of FRB hosts and the electron density field, which must be accounted for in cosmological analyses involving FRBs. In this section, we derive this effect by computing the angular cross-power spectrum between the electron distribution in the LSS and the spatial distribution of FRB host galaxies.

   We begin by expressing the total observed DM field in a given sky direction $\nv$ as an average over the DM of all FRBs along $\nv$ 
   \begin{align}\nonumber
     {\rm DM}_{\rm tot}(\nv)=\frac{1}{N_{\rm f}(\nv)}\int_0^\infty \mathrm{d}z\,n(z)\,[1+\delta_{\rm f}(\chi\nv,z)]\\\times\,\left({\rm DM}_{\rm LSS}(\nv,z)+\frac{{\rm DM}_{\rm host}(\nv,z)}{1+z}\right).
   \end{align}
   where ${\rm DM}_{\rm LSS}(\nv,z)$ is given in Eq. \eqref{eq:DM_LSS2}, and ${\rm DM}_{\rm host}(\nv,z)$ is the rest-frame host contribution for an FRB at redshift $z$ and sky position $\nv$. The quantity $N_{\rm f}(\nv)$ is the angular density of FRBs detected along $\nv$ and is given by
   \begin{equation}
     N_{\rm f}(\nv)\equiv\int_0^\infty \mathrm{d}z\,n(z)\,[1+\delta_{\rm f}(\chi\nv,z)].
   \end{equation}
   For simplicity, we have further assumed that the galactic contribution to the observed DM has been modelled and subtracted \citep{Yao2017new, Platts2020data}. We may now split ${\rm DM}_{\rm LSS}$ into its cosmological mean $\overline{\rm DM}_{\rm LSS}(z)$ and fluctuation ${\cal D}(\nv,z)$, as in Eq. \eqref{eq:DM_mean_fluct}. We can also express ${\rm DM}_{\rm host}$ as a sum of mean and spatial fluctuation, where the latter incorporates the halo-to-halo stochasticity in gas content for FRB-hosting haloes, as well as correlated fluctuations in the host DM contribution tracking the local LSS (we provide a toy example of such a contribution in Appendix \ref{app:host_more}). This leads to:
   \begin{equation}
     {\rm DM}_{\rm host}(\nv,z)=\overline{\rm DM}_{\rm host}(z)+{\cal D}_{\rm host}(\nv,z).
   \end{equation}
   Expanding $1/N_{\rm f}(\nv)$ to first order in $\delta_{\rm f}$, and keeping only terms linear in the perturbations ($\delta_{\rm f}$, ${\cal D}$, ${\cal D}_{\rm host}$), we find the total DM fluctuation to be:
   \begin{align}\label{eq:DM_with_host}
     {\cal D}_{\rm tot}(\nv)&={\cal D}(\nv)+\int \mathrm{d}z\,\frac{n(z)}{1+z}\,{\cal D}_{\rm host}(\nv,z)\\\nonumber
     &+\int \mathrm{d}z\,n(z)\left[\Delta\overline{\rm DM}_{\rm LSS}(z)+\Delta\left(\frac{\overline{\rm DM}_{\rm host}(z)}{1+z}\right)\right]\,\delta_{\rm f}(\chi\nv,z),
   \end{align}
   where we have defined
   \begin{equation}
     \Delta X(z)\equiv X(z)-\int \mathrm{d}z'\,n(z')\,X(z')
   \end{equation}
   as the difference of a given background quantity $X(z)$ with respect to its FRB-weighted average as a function of redshift. We thus see that the observed DM fluctuation receives additional contributions from the intrinsic fluctuation in the host DM, as given by the second term in the first line of Eq. \eqref{eq:DM_with_host}, and from the clustering of the FRBs coupled with the redshift evolution of the mean LSS and host DMs, as given in the second line of Eq. \eqref{eq:DM_with_host}.

   Here, we will use a relatively simple model. Firstly, we will neglect the intrinsic host fluctuation ${\cal D}_{\rm host}$ (or, in other words, assume that it is not significantly correlated with the LSS). Secondly, we will assume no evolution in the mean host contribution, which we take to be fixed at $\overline{\rm DM}_{\rm host} = 90 \, \mathrm{pc} \, \mathrm{cm}^{-3}$. Finally, we will model the FRB source overdensity as a linearly biased tracer of the matter fluctuations, $\delta_{\rm f}(\chi \nv)\simeq b_{\rm f}\,\delta_{\rm m}(\chi\nv)$, and we will treat the FRB bias, $b_{\rm f}$, as a free parameter. 

   \begin{figure*}
       \centering
       \includegraphics[width=0.9\linewidth]{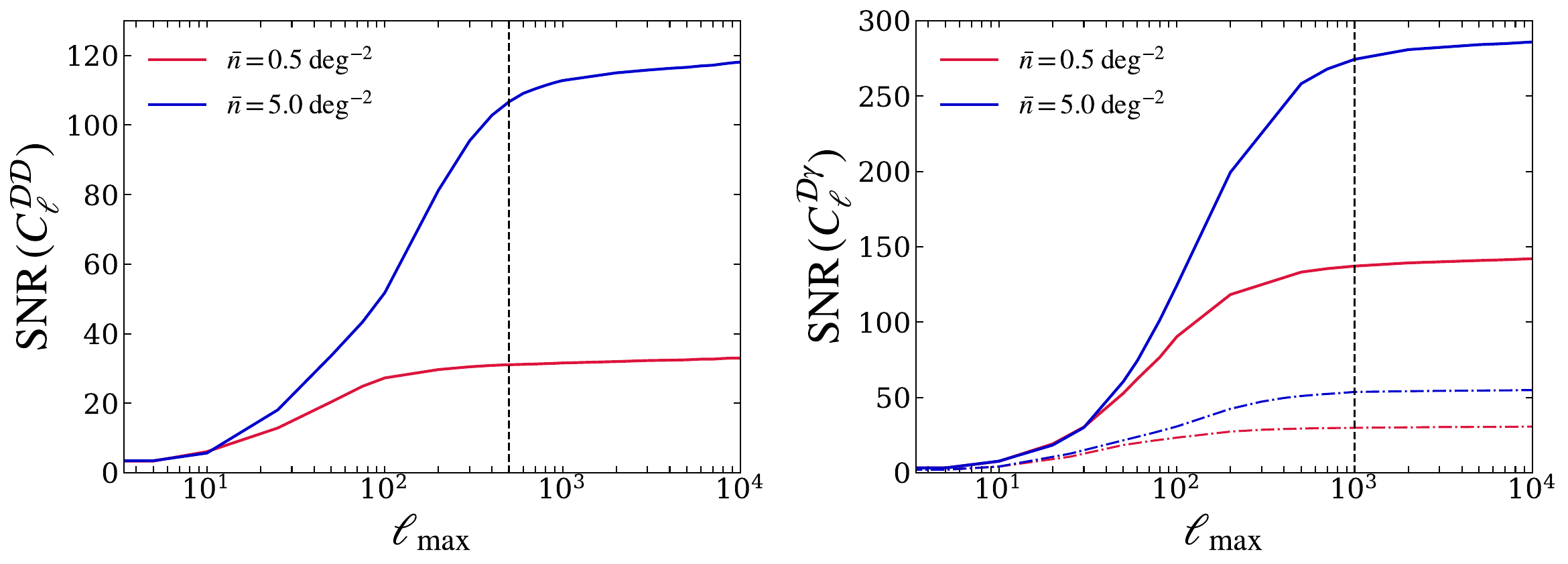}
       \caption{The signal-to-noise ratio, SNR, as a function of the maximum multipole, $\ell_{\rm max}$, for two FRB number densities. The left panel shows the SNR of the FRB auto-angular power spectrum, while the right panel shows the SNR of the FRB--WL cross-angular power spectrum for the first (dashed) and the final (solid) WL redshift bins. The red curves correspond to the fiducial case with $\bar{n} = 0.5 \; \mathrm{deg}^{-2}$, and the blue curves to a future survey with $\bar{n} = 5.0 \; \mathrm{deg}^{-2}$. The vertical dashed vertical lines indicate the adopted maximum scale cuts, $\ell^{{\cal D}{\cal D}}_{\rm max} = 500$ and $\ell^{{\cal D}\gamma}_{\rm max}=1000$.}
       \label{fig:snr_vs_ell_max}
   \end{figure*}
  
   Note that these simplifications are not overly prescriptive. Firstly, as shown in \cite{Alonso2021linear}, the FRB clustering contribution is likely relatively small (less than 10\% of the LSS signal). Hence, an approximate model to describe it is likely sufficient. Secondly, the neglected ${\cal D}_{\rm host}$ contribution has the same structure as the FRB clustering terms, with its radial kernel tracking the redshift distribution of FRBs in the sample. The impact from this term can therefore be absorbed approximately into the effective FRB bias parameter, $b_{\rm f}$, over which we marginalise.  It is important to note that we have dropped higher-order terms in deriving this expression. At such orders, the scale dependence of the host contribution becomes relevant, and accurately capturing this effect would require modelling both the spatial distribution of FRBs within their host haloes and the corresponding distribution of host DMs. In this work, however, we focus on the overall clustering amplitude, as the scale dependence enters only as a higher-order correction. Specifically, it arises as a bispectrum contribution, involving the three-point function of the local power spectrum and the matter overdensity, and is likely negligible in practise.

   To summarise, we include the contribution from source clustering as an additional projected tracer of the matter fluctuations $\delta_{\rm m}$ with a radial kernel given by
   \begin{equation}
     W_{\rm c}(\chi)=b_{\rm f}\,n(z)\,\left[\Delta\overline{\rm DM}_{\rm LSS}(z)+\overline{\rm DM}_{\rm host}\Delta\left(\frac{1}{1+z}\right)\right].
   \end{equation}
   The source-clustering term then introduces additional correlations between the DM fluctuations and any LSS tracer. In particular, the power spectra involving the probes considered here become
   \begin{align}
     C_{\ell}^{\mathcal{DD}} &= C_{\ell}^{\rm ee} + 2 C_{\ell}^{\rm ec} + C_{\ell}^{\rm cc}, \label{eq:Cl_DD} \\
     C_{\ell}^{\mathcal{D}\gamma} &= C_{\ell}^{\rm e\gamma} + C_{\ell}^{\rm c\gamma}, \label{eq:Cl_Dgamma} \\
     C_{\ell}^{\mathcal{D}\rm g} &= C_{\ell}^{\rm eg} + C_{\ell}^{\rm cg} \label{eq:Cl_Dg},
   \end{align}
   where the superscript ``$\rm c$'' denotes correlations with the source clustering term, and ``${\rm e}$'' denotes the standard LSS contribution to the DM (${\cal D}$). Fig. \ref{fig:cls_source_clustering} presents the individual contributions to the auto- and cross-power spectra involving DM correlations, as defined in Eqs. \eqref{eq:Cl_DD}-\eqref{eq:Cl_Dg}. The source clustering and large-scale structure power spectra exhibit distinct shapes, with the former having a significantly smaller amplitude for reasonable values of the FRB bias ($b_{\rm f} \in \{1.0, 3.0\}$).

  \subsubsection{FRB survey specifications} \label{sssec:model.frb.survey}
   We now specify the FRB survey characteristics and noise properties associated with the observed DMs. The DM signal receives contributions not only from the cosmological free-electron distribution, but also from the Milky Way and the host galaxies of FRBs. Since the Milky Way component acts as a foreground with a well-characterised spatial structure and is uncorrelated with the extragalactic LSS, its impact is expected to be most significant on large angular scales \citep{Yao2017new, Platts2020data}. We therefore treat it as a large-scale systematic and discuss its mitigation via scale cuts in Section \ref{sssec:results.systematics.mw}. On the other hand, the contribution from the FRB host galaxy introduces a stochastic variance $\sigma_{\rm host}^2$ due to intrinsic scatter among sources. This acts as a white noise term in the angular power spectrum, arising from the finite number of FRBs per unit solid angle, $\bar{n}$. Accounting for both the intrinsic DM variance, $\sigma_{\mathcal{D}}^2$, and the host contribution, the total noise power spectrum is given by
   \begin{equation} \label{eq:Nl_frb}
       N_\ell = \frac{\sigma_{\mathcal{D}}^2+\sigma_{\rm host}^2}{\bar{n}}
   \end{equation}
   where the intrinsic DM variance is
   \begin{equation}
       \sigma_{\mathcal{D}}^2 = \sum_\ell \frac{2\ell+1}{4\pi} \, C_\ell^{\mathcal{DD}},
   \end{equation}
   and the redshift-dependent host variance is
   \begin{equation} \label{eq:DM_sigma_host}
       \sigma_{\rm host}^2 = \int \mathrm{d}z \, n_{\rm FRB}(z) \frac{\sigma_{\rm host,0}^2}{(1+z)^2}.
   \end{equation}
   This expression arises from taking the average of the host variance over redshift, where the $1/(1+z)^2$ factor comes from translating from the rest frame of the FRB to that of the observer. The intrinsic DM variance, $\sigma_{\mathcal{D}}^2$, enters the noise term in Eq. \eqref{eq:Nl_frb} because the DM field is sampled only at discrete FRB positions, with self-pair contributions producing a scale-independent variance analogous to shot noise, while cross-pairs trace the angular correlations and contribute to the signal $C_\ell^{\mathcal{DD}}$ \citep{Wolz2025catalog}.

   We model the FRB redshift distribution, $n_{\rm FRB}(z)$, using a parametrisation commonly adopted in recent analyses \citep[e.g.][]{Reischke2023calibrating, Sharma2025probing}. Specifically, we assume
   \begin{equation} \label{eq:nz}
      n_{\rm FRB}(z) \propto (z+\delta z)^2 \exp(-\alpha z),
   \end{equation}
   where the parameter $\alpha$ governs the effective depth of the sample (i.e. the high-redshift tail of the distribution) and $\delta z$ allows for shifts in its mean redshift. This functional form captures the declining FRB detection efficiency at high redshift, driven by a combination of the FRB luminosity function, instrumentation selection effects, and intrinsic evolution of the underlying population. We assume that the redshifts are known for a representative subset of FRBs, while the overall redshift distribution remains subject to significant uncertainty. Consequently, we marginalise over $\alpha$ and $\delta z$ using broad priors, as listed in Table \ref{tab:free_parameters}.
   
   We consider the following survey configuration, based on current experiments such as CHIME/FRB \citep{Rafiei2021CHIME}. Specifically, we adopt $\alpha = 3.5$ and $\delta z = 0$ for the fiducial FRB redshift distribution, an FRB number density of $\bar{n} = 0.5 \, \mathrm{deg}^{-2}$, a host galaxy DM variance of $\sigma_{\rm host,0}=90\,{\rm pc \, cm}^{-3}$, and a survey area of $f_{\rm sky} = 0.7$. It is worth noting that the host galaxy DM variance plays a non-negligible role in parameter constraints. For example, increasing $\sigma_{\rm host,0}$ from $90\,{\rm pc \, cm}^{-3}$ to $180\,{\rm pc \, cm}^{-3}$ significantly degrades the precision on $\eta_{\rm b}$ by a factor of $\sim 2$, while only mildly affecting $\log_{10} M_{\rm c}$. This highlights the importance of accurately modelling this contribution \citep[e.g.][]{Reischke2023calibrating, Sharma2025probing}.

   For this survey configuration, we assess the detectability of the FRB auto-correlation and FRB--WL cross-correlation signals, and examine the improvement expected from a future survey with an order-of-magnitude increase in the FRB number density, $\bar{n} = 5.0 \, \mathrm{deg}^{-2}$. Such number densities may be achievable with a combination of next-generation instruments including SKA \citep{Lazio2009SKA}, DSA-2000 \citep{Hallinan2019dsa}, and CHORD \citep{Vanderlinde2020CHORD}, although in practise only a subset of detected FRBs will be sufficiently well-localised for clustering and cross-correlation analyses. 
   
   Fig. \ref{fig:snr_vs_ell_max} shows the signal-to-noise ratio, SNR, as a function of the maximum multipole, $\ell_{\rm max}$, for the FRB auto-angular power spectrum and the FRB--WL cross-angular power spectrum, considering the first and final WL redshift bins and illustrating the dependence on the FRB number density. Increasing the FRB number density enhances the SNR by a factor of $\sim 3$ for the auto-correlation and by $\sim 2$ for the cross-correlations. 
   
   For the cross-correlation with the final WL redshift bin, the covariance is dominated by cosmic variance, with a reduced contribution from shape noise. As a result, for a fixed FRB number density, the total SNR is enhanced by a factor of $\sim 5$ compared to the first WL redshift bin.

   Moreover, the SNR saturates at $\ell^{\cal{DD}} \gtrsim 500$ and $\ell^{\cal{D}\gamma} \gtrsim 1000$ for the auto- and cross-correlation analyses, respectively, indicating that small-scale modes contribute minimally to the total signal. Based on these results, we adopt $\ell_{\rm max}^{\cal{DD}} = 500$ and $\ell_{\rm max}^{\cal{D}\gamma} = 1000$ in our analysis.

  \subsection{Galaxy clustering} \label{ssec:model.galaxies}
   The overdensity in the distribution of galaxies is a biased tracer of the underlying matter overdensity. In the simplest model, galaxies are assumed to follow the matter field through a linear bias relation. However, this approximation holds on sufficiently large scales, whereas more sophisticated bias models are typically required at smaller scales \citep[see e.g.][for a review]{Desjacques2016bias}. 
   
   In this work, we investigate the potential of the cross-correlation between galaxies and FRBs to constrain baryonic effects. This requires probing sufficiently small scales to resolve the gas distribution within haloes. To model galaxy clustering on these scales, we adopt a halo occupation distribution (HOD) framework, as described below, and restrict our analysis to angular multipoles up to $\ell_{\rm max} = 1000$. 

   We consider two types of galaxy samples that probe haloes of different characteristic masses, specifically, Luminous Red Galaxies (LRGs) and Emission Line Galaxies (ELGs). For the LRG sample, we use the standard HOD framework \citep{Zheng2004theoretical, Ando2017angular, Nicola2020tomographic}. Under this model, the mean occupation numbers of central and satellite galaxies are given by
   \begin{equation}
       \left\langle N_{\rm cen}(M) \right\rangle = \frac{1}{2} \left[1+ \mathrm{erf}\left(\frac{\log(M/M_{\rm min})}{\sigma_{\ln M}}\right)\right]
   \end{equation}
   and
   \begin{equation} \label{eq:N_sat_hod}
       \left\langle N_{\rm sat}(M) \right\rangle = \Theta(M-M_0) \left(\frac{M-M_0}{M_1}\right)^\alpha,
   \end{equation}
   respectively. Here, $M_{\rm min}$ sets the threshold halo mass for central galaxy occupation, $\sigma_{\ln M}$ is the width of the mass distribution, $M_0$ is the minimum halo mass required for satellite occupation, $M_1$ is the characteristic mass scale at which haloes host on average one satellite galaxy, and $\alpha$ governs the increase in satellite richness as halo mass increases. The mean galaxy density profile is then given by
   \begin{equation}
       \left\langle n_{\rm g}(r)|M \right\rangle = \left\langle N_{\rm cen}(M) \right\rangle \left[f_{\rm c} + \left\langle N_{\rm sat}(M)\right\rangle u_{\rm sat}(r|M)\right],
   \end{equation}
   where $f_{\rm c}$ is the observed fraction of central galaxies and $u_{\rm sat}(r|M)$ is the radial distribution of satellite galaxies as a function of distance to the centre of the halo, which is assumed to follow an Navarro-Frenk-White (NFW) profile \citep{Navarro1996universal}. For our fiducial LRG sample, we adopt the following set of HOD parameters: $\{\log_{10} M_{\rm min}, \log_{10} M_0, \log_{10} M_1, \alpha, \sigma_{\ln M}\} = \{12.86, 12.47, 13.97, 1.27, 0.026\}$. These values are derived for the first redshift bin in \citet{Zhou2023desi} using a linear interpolator based on the HOD parameters reported in Table 4 of \citet{Zhou2020clustering}.
   
   For the ELG sample, we employ the star-forming halo occupation distribution (SFHOD) model, using the parametrisation and fits of \citet{Rocher2023desi}. The mean central occupation is given by
     \begin{equation}
      \left\langle N_{\rm cen}(M) \right\rangle = 
      \begin{cases}
          \left\langle N_{\rm cen}^{\rm GHOD}(M)\right\rangle & M \leq M_{\rm cen}, \\
          \frac{A_{\rm c}}{\sqrt{2\pi} \sigma_{\ln M}} \left(\frac{M}{M_{\rm min}}\right)^\gamma & M > M_{\rm cen},
      \end{cases}
  \end{equation}
  where the Gaussian HOD component is
  \begin{equation}
      \left\langle N_{\rm cen}^{\rm GHOD}(M)\right\rangle = \frac{A_{\rm c}}{\sqrt{2\pi} \sigma_{\ln M}} \exp\left(-\frac{(\log_{10} (M/M_{\rm min})^2}{2\sigma_{\ln M}^2}\right).
  \end{equation}
  Here, $A_{\rm c}$ controls the size of the central galaxy sample and $\gamma$ governs the high-mass asymmetry of the distribution. We adopt the following HOD for satellites in the ELG population:
  \begin{equation}
      \left\langle N_{\rm sat}(M) \right\rangle = A_{\rm s} \left(\frac{M-M_0}{M_1}\right)^\alpha,
  \end{equation}
  where $A_{\rm s}$ sets the size of the satellite galaxy sample, and the parameters $M_0$, $M_1$, and $\alpha$ are the same as defined in Eq. \eqref{eq:N_sat_hod}. For our fiducial ELG sample, we adopt the SFHOD fit reported in Table 2 of \citet{Rocher2023desi} for the DESI ELG One-Percent survey, i.e. $\{A_{\rm c}, A_{\rm s}, \log_{10} M_{\rm min}, \log_{10} M_0, \log_{10} M_1, \alpha, \sigma_{\ln M}, \gamma\} = \{1.0, 0.09, 11.87, 11.73, 9.30, -0.28, 0.07, -4.42\}$.

  Since our main aim is to explore the possibility of characterising the mass dependence of the gas model parameters through the cross-correlation of DM and galaxies of different types, we assume the same redshift distribution for the LRG and ELG samples for simplicity. Specifically, we adopt the first redshift bin of \citet{Zhou2023desi} for both galaxy populations. The resulting distribution has a mean redshift of 0.47 and a standard deviation of 0.06. The radial kernel for galaxy, $W_{\rm g}(\chi)$, is related to the redshift distribution, $n_{\rm g}(z)$, via
  \begin{equation} \label{eq:gc_kernel}
      W_{\rm g}(\chi) = \frac{H(z)}{c} \, n_{\rm g}(z).
  \end{equation}

 \subsection{Baryon model} \label{ssec:model.baryons}
  We incorporate baryonic effects within the halo model framework using the hydrostatic equilibrium model described in \citet{Ferreira2023xray} and \citet{LaPosta2025insights}. This implementation models the gas density profile as the sum of the bound and ejected gas, each described separately under the assumption of hydrostatic equilibrium within a dark matter halo. In the following subsections, we outline the adopted parametrisations for the mass fractions and density profiles of the relevant components. Specifically, the model employs the dark matter profile of \citet{Navarro1996universal}, the stellar profile of \citet{Fedeli2014clustering}, the bound gas profile of \citet{Mead2020hydrodynamical}, and the ejected gas profile of \citet{Schneider2015new}.
  
  \subsubsection{Dark matter}
   We consider haloes to be composed of cold dark matter (CDM), gas, and stars. The CDM mass fraction is fixed to the universal value
   \begin{equation}
       f_{\rm c}(M) = \frac{\Omega_{\rm c}}{\Omega_{\rm m}},
   \end{equation}
   which reflects the fact that baryonic feedback can redistribute CDM within haloes, but cannot eject it.

   Using gravity-only simulations, \citet{Navarro1996universal} demonstrated that dark matter haloes form approximately spherical structures described by the NFW profile:
   \begin{equation}
       \rho_{\rm c}(M,r) \propto \frac{1}{r/r_{\rm s}(1+r/r_{\rm s})^2},
   \end{equation}
   where $r_{\mathrm{s}}$ is the scale radius, which is related to the halo virial radius $r_{200\mathrm{c}}$ through $r_{\mathrm{s}} \equiv  r_{200\mathrm{c}}/ c$, with $c$ denoting the concentration parameter.

  \begin{table*}
      \centering
      \begin{tabular}{llrr} 
        \hline
        Parameter & Description & Fiducial Value & Prior \\
        \hline
        \textbf{Cosmological} &&& \\
        $\Omega_{\mathrm{m}}$ & density of matter in units of the critical density of the Universe & $0.30967$ & $\mathcal{N}(0.30967, 0.1)$ \\
        $\sigma_8$ & cold mass linear mass variance in $8 h^{-1} \text{Mpc}$ spheres & $0.8102$ & $\mathcal{N}(0.8102, 0.1)$ \\
        $\Omega_{\mathrm{b}}$ & density of baryons in units of the critical density of the Universe & $0.04897$ & $-$ \\
        $h$ & dimensionless Hubble constant & $0.6766$ & $\mathcal{N}(0.6766, 0.1)$ \\
        $n_{\mathrm{s}}$ & scalar spectral index & $0.9665$ & $\mathcal{N}(0.9665, 0.1)$ \\
        $\Sigma m_\nu$ & sum of neutrino masses in units of eV & $0.06$ & $\mathcal{N}(0.06, 0.1)$ \\ 
        \hline
        \textbf{Baryonic Feedback} &&& \\
        $\log_{10} M_{\mathrm{c}}$ & characteristic halo mass for which half the gas is retained & $14.0$ & $\mathcal{N}(14.0, 5.0)$ \\
        $\eta_{\rm b}$ & directly proportional to the radius of ejected gas from the halo & $0.5$ & $\mathcal{N}(0.5, 5.0)$ \\
        $\beta$ & describes the rate at which the depletion of gas increases towards smaller haloes & $0.6$ & $-$ \\
        $\gamma$ & related to the polytropic index of the gas via $\gamma = 1/(\Gamma-1)$ & $1.17$ & $-$ \\
        $A_*$ & controls the stellar abundance & $0.03$ & $-$ \\
        \hline
        \textbf{Intrinsic Alignments} &&& \\
        $A_{\mathrm{IA},0}$ & amplitude of the NLA model for IAs & $0.0$ & $\mathcal{N}(0, 1.0)$\\
        $\eta_{\mathrm{IA}}$ & slope of the NLA model for IAs & $0.0$ & $\mathcal{N}(0, 1.0)$\\
        \hline
        \textbf{Fast Radio Bursts} &&& \\
        $\sigma_{\rm host,0}$ & host galaxy DM variance & 90 $\mathrm{pc} \, \mathrm{cm}^{-3}$ & $-$ \\
        $\bar{n}$ & FRB number density & 0.5 $\mathrm{deg}^{-2}$ & $-$ \\
        $\alpha$ & effective depth of FRB sample & 3.5 & $\mathcal{N}(3.5, 1.0)$ \\
        $\delta z$ & shift parameter for the mean redshift of the FRB distribution & 0.0 & $\mathcal{N}(0, 1.0)$ \\
        $b_{\rm f}$ & FRB bias, a constant relating FRB overdensity to the matter overdensity & 1.0 & $\mathcal{N}(1.0, 1.0)$ \\
        \hline
        \textbf{LRG Sample} &&& \\
        $\log_{10} M_{\rm min}$ & threshold halo mass for central galaxy occupation & 12.86 & $\mathcal{N}(12.86, 1.0)$ \\
        $\log_{10} M_1$ & characteristic mass at which haloes host on average one satellite galaxy & 13.96 & $\mathcal{N}(13.96, 1.0)$ \\
        $\log_{10} M_0$ & minimum halo mass required to host satellite galaxies & 12.47 & $-$ \\
        $\sigma_{\ln M}$ & sets the width of the mass distribution & 0.026 & $-$ \\
        $\alpha$ & satellite power-law index & 1.27 & $-$ \\
        \hline
        \textbf{ELG Sample} &&& \\
        $\log_{10} M_{\rm min}$ & threshold halo mass for central galaxy occupation & 11.87 & $\mathcal{N}(11.87, 1.0)$ \\
        $\log_{10} M_1$ & characteristic mass at which haloes host on average one satellite galaxy & 9.30 & $\mathcal{N}(9.30, 1.0)$ \\
        $\log_{10} M_0$ & minimum halo mass required to host satellite galaxies & 11.73 & $-$ \\
        $\sigma_{\ln M}$ & sets the width of the mass distribution & 0.07 & $-$ \\
        $\alpha$ & satellite power-law index & $-0.28$ & $-$ \\
        $A_{\rm c}$ & controls the size of the central galaxy sample & 1.0 & $-$ \\
        $A_{\rm s}$ & controls the size of the satellite galaxy sample & 0.09 & $-$ \\
        $\gamma$ & controls the high-mass asymmetry of the mass distribution & $-4.42$ & $-$ \\
        \hline
      \end{tabular}
      \caption{Qualitative descriptions of cosmological parameters, baryonic parameters used in the hydrostatic equilibrium model, intrinsic alignment parameters, FRB survey configuration parameters, and HOD parameters for the LRG and ELG samples. Note that all masses are in units of $M_{\odot}$ throughout this work. Here, fiducial value refers to the parameter value used to generate the mock data. We use Gaussian priors, centred on the corresponding fiducial value, for all free parameters.}
      \label{tab:free_parameters}
    \end{table*}
   
  \subsubsection{Central galaxy}
   The stellar mass fraction follows the parametrisation proposed by \citet{Fedeli2014clustering}:
   \begin{equation} \label{eq:f_star}
       f_*(M) = A_* \, \exp\left[-\frac{1}{2}\left(\frac{\log_{10}(M/M_*)}{\sigma_*}\right)^2\right],
   \end{equation}
   where $M_* = 10^{12.5} \, M_\odot/h$, $\sigma_* = 1.2$, and $A_* = 0.03$ \citep{Moster2013galactic, Kravtsov2014stellar}. This Gaussian form encapsulates the trend that star formation efficiency peaks in haloes of mass $M_*$ with halo stellar mass fraction $A_*$, and declines at both higher and lower halo masses with a logarithmic width $\sigma_*$. Given that the stellar component contributes only at very small distances from the central galaxy, we model its scale-dependent profile as a Dirac delta at $r=0$.

  \subsubsection{Bound gas}
   The total gas budget in a halo can be decomposed into a component gravitationally bound within haloes and a component ejected beyond the virial radius by baryonic feedback processes. The bound gas mass fraction is modelled as
   \begin{equation} \label{eq:f_bgas}
       f_{\rm b}(M) = \frac{\Omega_{\rm b}/\Omega_{\rm m} - f_*}{1+(M_{\rm c}/M)^\beta},
   \end{equation}
   where $\Omega_{\rm b}$ is the cosmic baryon density with respect to the critical density, $M_{\rm c}$ denotes the characteristic mass scale at which half of the gas has been ejected from the halo by AGN-driven outlows, and $\beta$ controls the steepness of the depletion with halo mass.

   For the spatial distribution of bound gas, we adopt the profile of \citet{Martizzi2013cusp}, which is a simplified version of the model from \citet{Komatsu2001universal}, chosen for its simplicity for calculations in Fourier space. The profile is parametrised as
   \begin{equation}
       \rho_{\rm b}(M,r) \propto \left[\frac{\ln(1+r/r_{\rm s})}{r/r_{\rm s}}\right]^{1/(\Gamma-1)},
   \end{equation}
   where $\Gamma$ is the gas polytropic index, which we fix to $\Gamma = 1.17$.
   
  \subsubsection{Ejected gas}
   The ejected gas fraction accounts for the remaining baryons originally associated with the halo. It is simply given by
   \begin{equation} \label{eq:f_egas}
       f_{\rm e}(M) = \Omega_{\rm b}/\Omega_{\rm m} - f_*(M) - f_{\rm b}(M),
   \end{equation}
   where $\Omega_{\rm b}/\Omega_{\rm m}$ is the total baryon mass fraction, $f_*(M)$ is the stellar mass fraction given by Eq. \eqref{eq:f_star}, and $f_{\rm b}(M)$ is the bound gas mass fraction from Eq. \eqref{eq:f_bgas}.

   The ejected gas density profile follows the Gaussian model of \citet{Schneider2015new}, which is motivated by the assumption that gas particles are expelled with a Maxwellian velocity distribution:
   \begin{equation}
       \rho_{\rm e}(r,M) \propto \frac{1}{(2\pi r_{\rm ej}^2)^{3/2}} \exp\left(-\frac{r^2}{2r_{\rm ej}^2}\right),
   \end{equation}
   where the characteristic scale $r_{\rm ej}$ corresponds to the mean distance travelled by a gas particle moving at the escape velocity. Following \citet{Schneider2015new}, we parametrise this scale as
   \begin{equation}
       r_{\rm ej} = 0.375 \, \sqrt{\Delta} \, \eta_{\rm b} r_\Delta,
   \end{equation}
   where $\Delta=200$ and $\eta_{\rm b}$ quantifies the distance to which gas is expelled by the AGN.

 \subsection{Likelihood} \label{ssec:model.likelihood}
  Throughout this work, we adopt a {\sl Planck} cosmology \citep{Planck2020} as our fiducial model, with parameter values $\{\Omega_{\mathrm{c}}, \Omega_{\mathrm{b}}, h, n_{\mathrm{s}}, \sigma_8, \Sigma m_{\nu}\} =$ $\{0.2607$, $0.04897$, $0.6766$, $0.9665$, $0.8102, 0.06\}$. This cosmology is used to generate the mock data vectors, $\mathbf{d}$, for the observables. We include multipoles up to $\ell_{\rm max} = \{500, 1000, 2000\}$ for the FRB auto-correlations, WL--FRB cross-correlations, and WL auto- and cross-correlations, respectively, and $\ell_{\rm max} = 1000$ for correlations involving galaxy clustering. The choices of $\ell_{\rm max}$ for correlations involving FRBs are motivated by the angular resolution of CHIME/FRB, which is approximately $0.2^\circ$ (corresponding to $\ell \sim 1000$) \citep{Rafiei2021CHIME}. Multipoles above this scale cannot be reliably measured, so these scale cuts are applied to ensure that the analysis remains robust while retaining the majority of the measurable signal.
  
  We adopt survey sky fractions of $f_{\rm sky}=\{0.4, 0.7, 0.35\}$ for the mock WL, FRB, and GC data, respectively. The slightly smaller fraction used for galaxy clustering accounts for the more stringent requirements imposed on the final survey geometry by the need to select a sufficiently homogeneous sample of sources. When computing the cross-correlation of two tracers, we take their overlapping sky area, corresponding to their minimum sky fraction.

  The covariance matrix of the data is constructed following the Knox formula for Gaussian fields \citep{Knox1996cosmic}:
  \begin{equation} \label{eq:knox}
     \mathsf{C}(C_{\ell}^{ij}, C_{\ell'}^{kl}) = \frac{\delta_{\ell \ell'}}{(2\ell+1) f_{\rm sky}} \left(\tilde{C}_{\ell}^{ik} \tilde{C}_{\ell}^{jl} + \tilde{C}_{\ell}^{il} \tilde{C}_{\ell}^{jk}\right),
  \end{equation}
  where
  \begin{equation}
      \tilde{C}_{\ell}^{ij} = C_{\ell}^{ij} + N_{\ell}^i \,\delta_{ij}.
  \end{equation}
  Here, $N_{\ell}^i$ denotes the noise contribution for tracer $i$ (i.e. DM/host variance contribution for FRB, intrinsic shape noise for weak lensing, and shot noise for galaxy clustering. The indices $\{i,j,k,l\}$ run over the set of tracers $\{\mathcal{D}, \gamma_{\rm a}, \mathrm{g}_{\rm b}\}$, corresponding to FRB DMs, cosmic shear in tomographic bin $\rm a$, and galaxy clustering in tomographic bin~$\rm b$. Assuming a Gaussian likelihood, the $\chi^2$ statistic for a parameter set $\boldsymbol{\theta}$ is then defined as
  \begin{equation}
      \chi^2 = - 2 \log p(\mathbf{d}|\boldsymbol{\theta}) = (\mathbf{d} - \mathbf{t}(\boldsymbol{\theta}))^{\mathrm{T}} \tilde{\mathsf{C}}^{-1} (\mathbf{d} - \mathbf{t}(\boldsymbol{\theta})) + K,
  \end{equation}
  where $\mathbf{t}$ denotes the theoretical prediction for the data $\mathbf{d}$, and $K$ is a normalisation constant. We sample the posterior distribution using the \texttt{Cobaya} framework \citep{Torrardo2021cobaya}, which implements the Metropolis-Hastings Markov Chain Monte Carlo (MCMC) algorithm \citep{Metropolis1953equation}. In our analysis, we sample over five cosmological parameters $(\Omega_{\mathrm{m}}, \sigma_8, h, n_{\mathrm{s}}, \Sigma m_{\nu})$, two WL IA parameters $(A_{\mathrm{IA,0}}, \eta_{\mathrm{IA}})$ as defined in Eq. \eqref{eq:NLA_IA}, and two baryonic feedback parameters $(\log_{10} M_{\mathrm{c}}$, $\eta_{\rm b})$. We fix the cosmological baryon fraction, $\Omega_{\rm b}$, to its fiducial value, given its precise determination from both the Cosmic Microwave Background (CMB) and Big Bang Nucleosynthesis (BBN). Additionally, we marginalise over the calibratable systematics affecting cosmic shear (i.e. photometric redshift uncertainties and multiplicative shape biases) using the analytical approximation described in Section \ref{sssec:model.shear.systematics}. In analyses incorporating FRB DMs, we marginalise over the relevant FRB nuisance parameters, specifically the redshift parameters $\alpha$ and $\delta z$, and the FRB clustering bias, $b_{\rm f}$. For analyses involving GC, we additionally marginalise over the HOD parameters $\log_{10} M_{\rm min}$ and $\log_{10} M_1$. A summary of all free parameters and their priors is presented in Table \ref{tab:free_parameters}.

  It is worth noting that, in addition to the DM clustering statistics used here, in principle, the variance of the FRB DM--redshift relation provides additional information that could be used to enhance constraints on cosmological and baryonic parameters \citep[see e.g.][]{Reischke2025first, Sharma2025hydrodynamical, Medlock2025constraining}. Since the exact form of the likelihood needed to incorporate this information is not trivial, this is left for future work.

\section{Results} \label{sec:results}
 \subsection{FRBs as an external calibrator of baryonic effects} \label{ssec:results.ext}
  In this section, we consider FRBs as an external calibrator of baryonic effects for WL analyses. We begin by quantifying the extent to which having freedom over the baryonic feedback parameters degrades the cosmological constraints for LSST-like cosmic shear data. We then perform a full 3$\times$2-point analysis that combines the mock WL measurements with the DMs of the mock FRB sample, calculating all combinations of power spectra, $\{C_\ell^{\mathcal{DD}}, C_\ell^{\mathcal{D}\gamma}, C_\ell^{\gamma\gamma}\}$.

  \begin{figure}
    \centering
    \includegraphics[width=\linewidth]{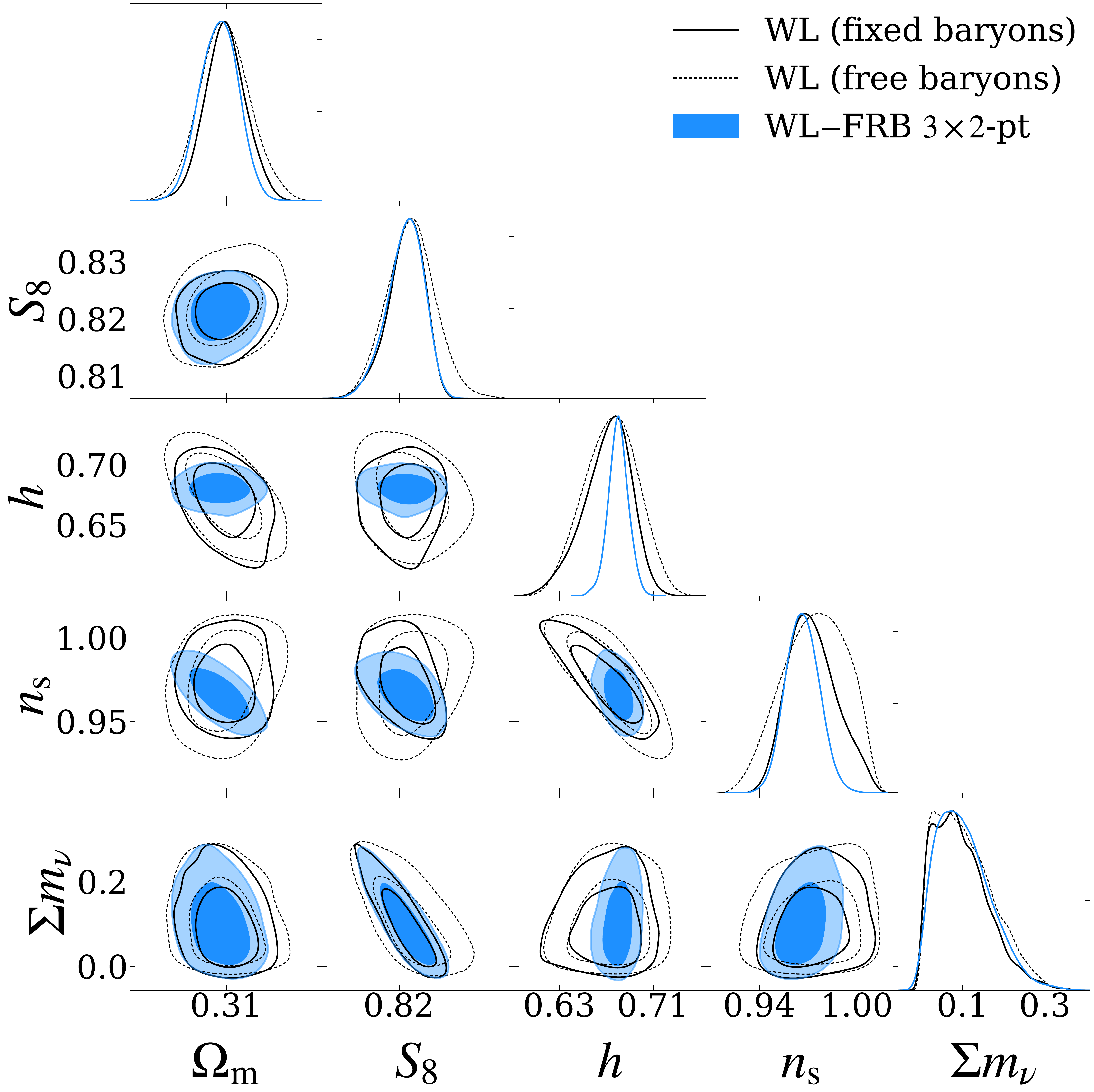}
    \caption{The marginalised posteriors on cosmological parameters for LSST-like weak lensing data only for fixed baryonic parameters (solid black), LSST-like weak lensing data only with marginalisation over baryonic parameters (dashed black), and a joint analysis of LSST-like weak lensing data with FRB DM correlations in a 3$\times$2-point analysis (blue). The inner and outer contours show the 68 per cent and 95 per cent confidence levels, respectively. We marginalise over intrinsic alignments, photometric redshift uncertainties, and multiplicative shape biases for the weak lensing data. We include multipoles up to $\ell_{\rm max} = \{500, 1000, 2000\}$ for the FRB auto-correlations, WL--FRB cross-correlations, and WL auto- and cross-correlations, respectively.}
    \label{fig:ext_calibrator}
  \end{figure}

  \begin{figure*}
    \centering
    \includegraphics[width=\linewidth]{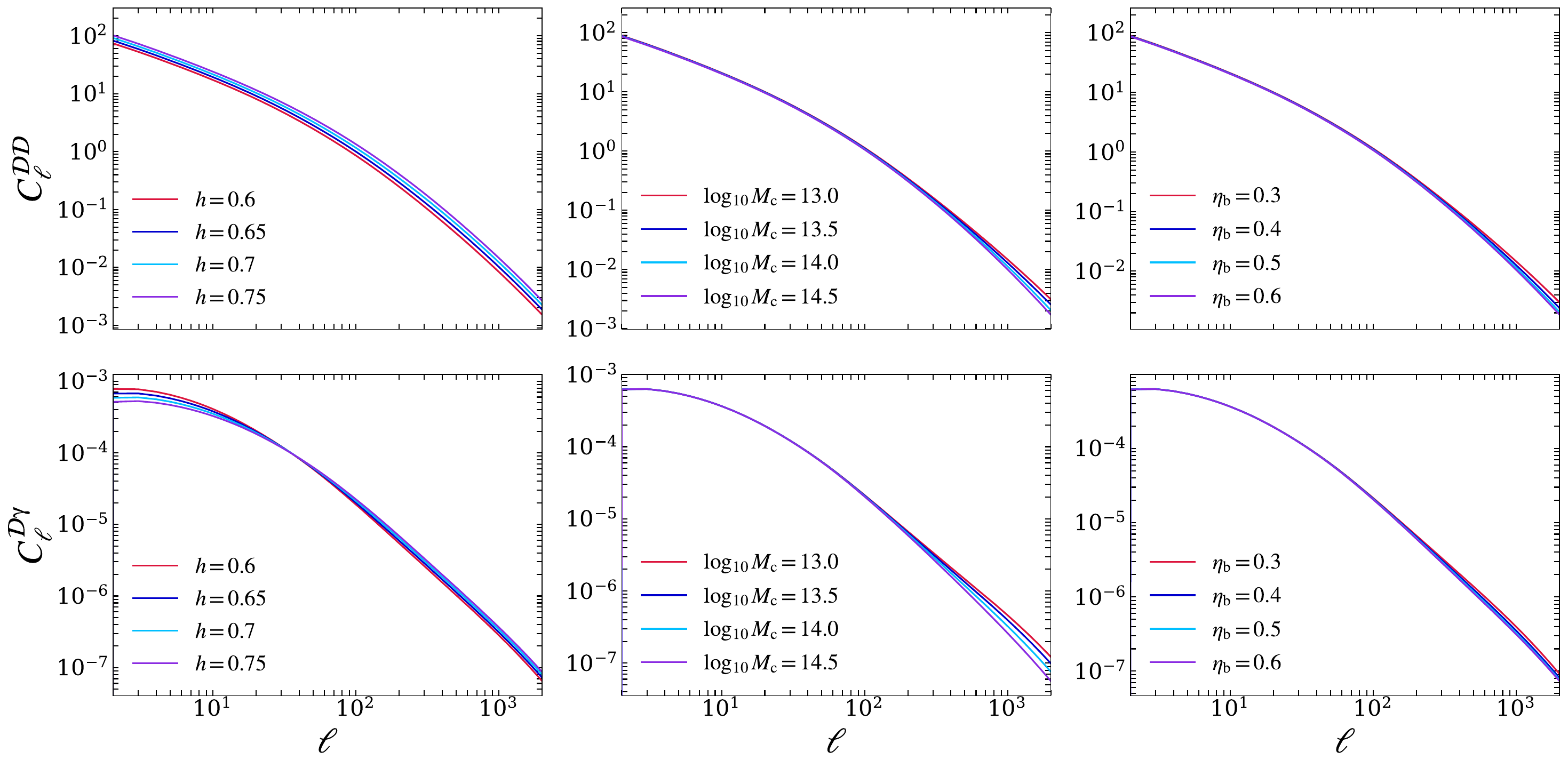}
    \caption{The impact of varying the parameters $h$, $\log_{10} M_{\rm c}$, and $\eta_{\rm b}$ on the $C_{\ell}^{\mathcal{DD}}$ (top) and $C_{\ell}^{\mathcal{D}\gamma}$ (bottom) angular power spectra. Each column isolates the response to a single parameter by varying it around the fiducial model, while keeping the remaining parameters fixed at their fiducial values.}
    \label{fig:h_sensitivity}
  \end{figure*}

  Fig. \ref{fig:ext_calibrator} compares the posterior constraints obtained from WL alone assuming perfect knowledge of baryonic effects (solid black) with those obtained after marginalising over baryonic feedback parameters (dashed black). The blue contours show the marginalised posteriors from the combined 3$\times$2-point analysis of WL and FRB DMs, under the idealised assumption that the FRB redshift distribution and source clustering contributions are perfectly known (we will examine the impact of these systematics in the following section). For WL alone, we find that the error in $S_8 \equiv \sigma_8 \sqrt{\Omega_{\rm m}/0.3}$, the parameter best determined by current weak lensing surveys, increases by a factor of 1.4 when marginalising over baryonic parameters. We find that the joint WL-FRB 3$\times$2-point analysis offers a significant improvement in cosmological constraints compared to WL alone, with the degradation factor on $S_8$ decreasing from 1.4 to 1.0.

  In addition to the improvements in the $(\Omega_{\rm m}, \sigma_8)$ plane, we find that the inclusion of FRB DMs leads to a notable tightening of constraints on the Hubble parameter, $h$. In standard WL analyses with fixed baryonic physics, $h$ is only weakly constrained because the shear power spectra are largely insensitive to the absolute distance scale, with any residual sensitivity arising primarily through the dependence of the matter power spectrum shape on the horizon scale at matter-radiation equality \citep[i.e. the Meszaros effect][]{Meszaros1976behaviour}. When FRBs are included, however, the uncertainty on $h$ decreases significantly, which indicates that the improvement is driven directly by the FRB measurements. This behaviour can be understood as follows. The DM can be interpreted as an integral constraint on the distance-redshift relation, with $\mathrm{DM} \propto h$ for a fixed baryon density. As a result, provided that the FRB redshift distribution is well known, the FRB auto-correlation and its cross-correlation with WL carry direct information about the expansion rate, allowing the joint WL--FRB analysis to break the degeneracy and substantially improve constraints on $h$.
  
  We investigate this behaviour by inspecting the sensitivity of the FRB--FRB and WL--FRB angular power spectra to variations in $h$, compared with variations in the main baryonic feedback parameters, $\log_{10} M_{\rm c}$ and $\eta_{\rm b}$. Fig. \ref{fig:h_sensitivity} illustrates that the dependence on $h$ of these power spectra is distinct from the parameters governing baryonic suppression, providing insight into how the FRB contribution sharpens the posterior for $h$. Specifically, we observe that both $\log_{10} M_{\rm c}$ and $\eta_{\rm b}$ influence the small-scale slope of the FRB--FRB power spectrum. We observe a similar behaviour in the case of $C_\ell^{{\cal D}\gamma}$, although the impact of $\log_{10}M_{\rm c}$ seems more pronounced in that case. This behaviour makes sense: the baryonic parameters govern the distribution of gas (the bound fraction and the spatial extent of the ejected gas) within each halo, but they do not change the overall abundance of gas associated with the halo, with the total baryonic fraction fixed to the cosmic average $\Omega_{\rm b}/\Omega_{\rm m}$. Thus, varying $\log_{10}M_{\rm c}$ or $\eta_{\rm b}$ can only modify the clustering on sub-halo scales, whereas on large scales, the two-halo regime is left unchanged. In turn, we see that varying $h$ modifies both power spectra on all scales. The scaling of the redshift-distance and redshift-DM relations with this parameter, leads to an overall scaling of the FRB auto-correlation, and a change in the broadband slope in the WL--FRB cross-correlation.

  Since the inclusion of additional cosmological degrees of freedom does not qualitatively modify our conclusions for the main parameters of interest, we adopt a simplified cosmological model for the remainder of the analysis in which only $\Omega_{\rm m}$ and $\sigma_8$ are treated as free. The extended cosmological parameter space yields results that are consistent with the simplified case, with the main exception of $h$, for which we have highlighted the distinctive improvement enabled by the FRB data. For clarity, we therefore focus on the $\{\Omega_{\rm m}, \sigma_8\}$ constraints in our fiducial analysis. In this case, we find that the error in $S_8$ increases by a factor of 2.2 when marginalising over baryonic parameters for WL alone, with the joint WL--FRB 3$\times$2-point analysis decreasing the degradation factor on $S_8$ to 1.2. The numerical constraints are listed in the first three rows of Table \ref{tab:cosmo_errors}.

  \subsubsection{Assessing the sensitivity of FRBs to baryonic effects}
   As a further step, we compare the constraining power of FRBs on baryonic physics with that of other established baryonic tracers. Specifically, we consider mock measurements of the X-ray-derived bound gas fraction and the stacked kSZ temperature profile, adopting the long-term datasets presented in \citet{Wayland2025calibrating}. Specifically, the X-ray dataset is constructed from a sample of 5259 galaxy groups and clusters detected by the eROSITA survey \citep{Ghirardini2024eROSITA}, while the kSZ dataset assumes the sensitivity and sky coverage of a CMB-S4-like experiment \citep{Schiappucci2025constraining}. As demonstrated in previous studies \citep[e.g.][]{Bigwood2024weak, Ferreira2023xray, Kovac2025baryonification, LaPosta2025insights, Wayland2025calibrating}, these probes are particularly effective at constraining the bound and ejected gas components of haloes, respectively.

   Comparing these tracers with FRBs therefore provides a complementary view of baryonic processes across different spatial scales. While X-ray and kSZ measurements are primarily sensitive to gas associated with haloes, FRBs probe the total line-of-sight electron content and thus offer sensitivity to more diffuse baryons. This comparison enables a direct assessment of the relative constraining power of each tracer, highlighting the potential benefits of combining them to break parameter degeneracies and obtain a more complete picture of baryonic feedback.
  
\begin{figure}
     \centering
     \includegraphics[width=\linewidth]{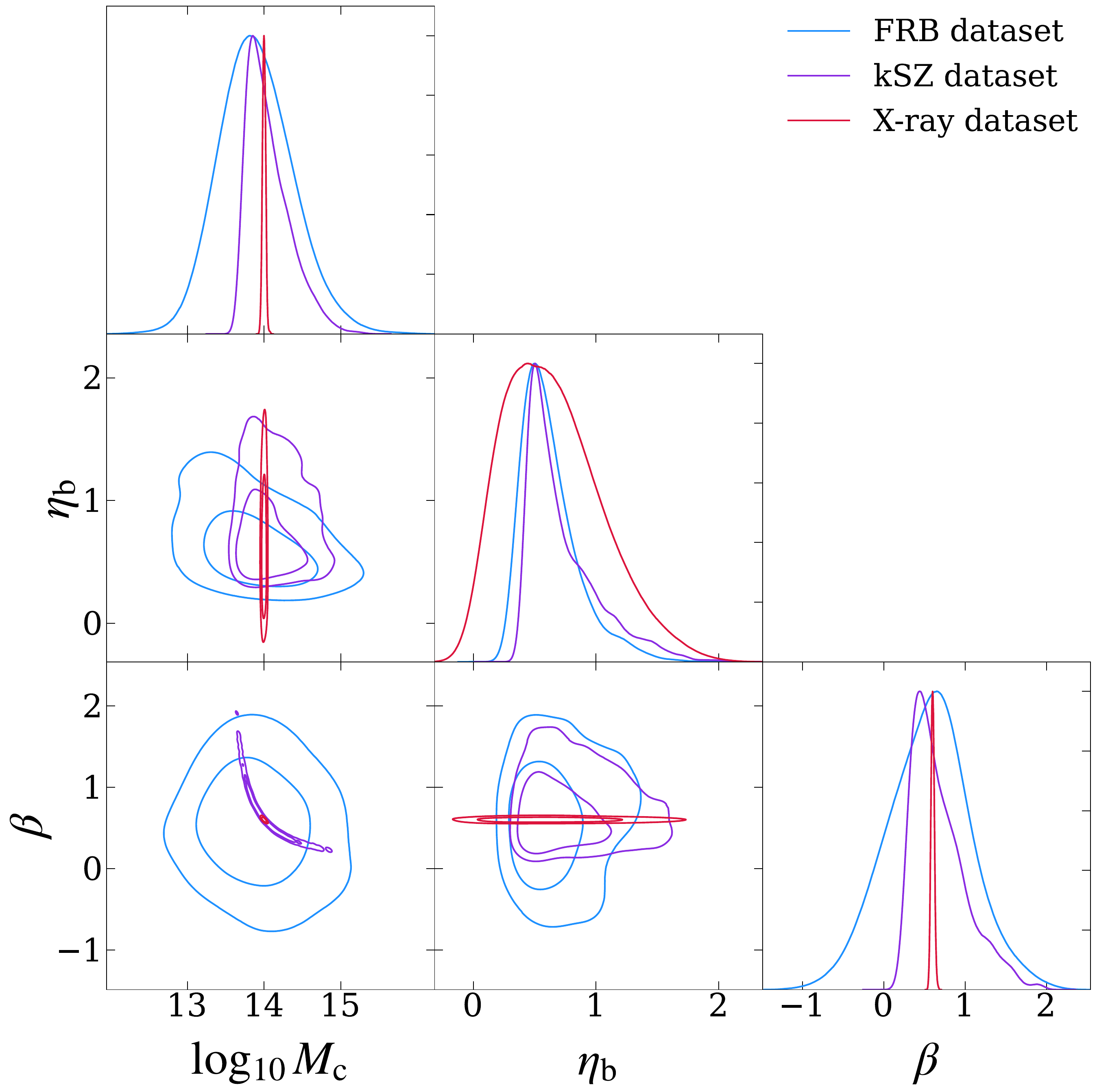}
     \caption{The marginalised posteriors on $\log_{10} M_{\rm c}$ and $\eta_{\rm b}$ for three different probes of baryonic physics with fixed cosmological parameters. The blue, purple, and red contours correspond to the constraints obtained from FRB DMs alone, kSZ alone, and X-rays alone, respectively. The inner and outer contours show the 68 per cent and 95 per cent confidence levels, respectively. The mock measurements of X-ray gas fractions are based on a sample of 5259 clusters from the eROSITA dataset, and the mock measurements of the stacked kSZ profile assume a CMB-S4-like experiment.}
     \label{fig:comparing_tracers}
   \end{figure}

   To investigate this, we fix all cosmological parameters to their fiducial values and vary only the baryonic parameters $\log_{10} M_{\rm c}$, $\eta_{\rm b}$, and $\beta$. The resulting constraints are shown in Fig. \ref{fig:comparing_tracers}, where the blue, purple, and red contours correspond to the FRB, kSZ, and X-ray measurements, respectively. We find that FRB data alone exhibits a strong degeneracy between $\log_{10} M_{\rm c}$ and $\eta_{\rm b}$, preventing tight constraints on either parameter. This is further illustrated in Fig. \ref{fig:h_sensitivity}, which shows that independent variations of $\log_{10} M_{\rm c}$ and $\eta_{\rm b}$ produce similar changes in the power spectrum on scales sensitive to baryonic physics. 
  
   Similarly, kSZ measurements alone are unable to place stringent constraints on $\log_{10} M_{\rm c}$ due to a degeneracy with $\beta$. As discussed in \citet{Wayland2025calibrating}, this degeneracy may be broken by including additional kSZ observations at different redshifts and stellar masses. In contrast, X-ray observations tightly constrain both $\log_{10} M_{\rm c}$ and $\beta$, as they directly probe the bound gas component within haloes. Quantitatively, X-ray data improves the uncertainties on $\log_{10} M_{\rm c}$ and $\beta$ by factors of 23.2 and 25.9, respectively, relative to the FRB-only case.
  
   Interestingly, neither X-ray nor kSZ data significantly improve constraints on $\eta_{\rm b}$ relative to those obtained from FRBs. In fact, for X-ray observations, the uncertainty on $\eta_{\rm b}$ increases by a factor of 1.7 relative to the FRB-only dataset, reflecting the limited sensitivity of X-rays to the ejected gas component of haloes. In contrast, kSZ measurements offer a slight improvement in the constraint on $\eta_{\rm b}$, reducing the uncertainty by a factor of 1.1 compared to FRBs alone. 
   
   It is important to highlight that these results are optimistic in the sense that we have not marginalised over the systematics affecting each tracer. While the impact of some systematics affecting the mock X-ray and kSZ measurements is explored in \citet{Wayland2025calibrating}, we defer a detailed treatment of FRB-related systematics to the following section.

  \begin{table*}
    \centering
    \begin{tabular}{lccc}
        \hline
        Tracer(s) & $\Omega_{\mathrm{m}}$ & $S_8$ & Degradation Factor on $S_8$ \\
        \hline
        WL only with baryons fixed & $0.3098^{+0.0039}_{-0.0043}$ & $0.8230^{+0.0025}_{-0.0024}$ & $-$ \\
        & & & \\
        WL only & $0.3099^{+0.0054}_{-0.0050}$ & $0.8238^{+0.0056}_{-0.0053}$ & 2.2 \\
        & & & \\
        WL + FRBs & $0.3092^{+0.0041}_{-0.0038}$ & $0.8226^{+0.0030}_{-0.0030}$ & 1.2 \\
        & & & \\
        WL + FRBs with systematics & $0.3097^{+0.0042}_{-0.0039}$ & $0.8233^{+0.0032}_{-0.0032}$ & 1.3 \\
        & & & \\
        WL + long-term FRBs & $0.3097^{+0.0033}_{-0.0033}$ & $0.8225^{+0.0026}_{-0.0024}$ & 1.0 \\
        \hline 
        WL + GC only with baryons fixed (LRG only) & $0.3096^{+0.0038}_{-0.0037}$ & $0.8231^{+0.0022}_{-0.0022}$ & $-$ \\
        & & & \\
        WL + GC only (LRG only) & $0.3100^{+0.0052}_{-0.0050}$ & $0.8237^{+0.0053}_{-0.0051}$ & 2.4 \\
        & & & \\
        WL + FRBs + GC (LRG only) & $0.3093^{+0.0037}_{-0.0036}$ & $0.8229^{+0.0031}_{-0.0030}$ & 1.4 \\
        \hline
        WL + GC only with baryons fixed (LRG + ELG) & $0.3097^{+0.0036}_{-0.0033}$ & $0.8231^{+0.0020}_{-0.0022}$ & $-$ \\
        & & & \\
        WL + GC only (LRG + ELG) & $0.3099^{+0.0043}_{-0.0039}$ & $0.8238^{+0.0052}_{-0.0050}$ & 2.4 \\
        & & & \\
        WL + FRBs + GC (LRG + ELG) & $0.3095^{+0.0033}_{-0.0034}$ & $0.8227^{+0.0030}_{-0.0030}$ & 1.4 \\
        \hline
    \end{tabular}
    \caption{The constraints on the cosmological parameters $\Omega_{\mathrm{m}}$ and $S_8 \equiv \sigma_8 \sqrt{\Omega_{\rm m}/0.3}$ obtained from different combinations of large-scale structure tracers. We report the mean marginal value of each parameter and their associated errors given by the 95$\%$ confidence level. Here, WL refers to LSST-like weak lensing data. We marginalise over intrinsic alignments, photometric redshift uncertainties, and multiplicative shape biases in the WL data. In all cases other than the first row, we marginalise over the parameters $\log_{10} M_{\rm c}$ and $\eta_{\rm b}$ describing baryonic effects. In the final column, we present the degradation factor on $S_8$, defined as the ratio of the error on $S_8$ for the tracer under consideration to that for the WL (or WL + GC) only fixed baryons case.}
    \label{tab:cosmo_errors}
  \end{table*}
  
 \subsection{Impact of FRB systematics} \label{ssec:results.systematics}
  Here, we perform a 3$\times$2-point analysis of WL and FRBs in which we marginalise simultaneously over all FRB systematics considered in this work, specifically, the redshift parameters $(\alpha, \delta z)$ and the FRB bias $(b_{\rm f})$. This provides a more realistic treatment of the mock FRB sample compared to the optimistic case presented in Section \ref{ssec:results.ext}. In this framework, we find that the WL--FRB 3$\times$2-point analysis is unable to place tight constraints on these nuisance parameters, which therefore remain prior-dominated.

  For unlocalised FRBs, the source redshift cannot be uniquely inferred from the DM due to its degeneracy with the line-of-sight electron density and source distance. To account for the limited knowledge of the true redshift distribution, we adopt broad priors on the nuisance parameters $\alpha$ and $\delta z$ (as listed in Table \ref{tab:free_parameters}) and marginalise over them in our analysis. We find that the resulting posteriors on $\alpha$ and $\delta z$ are largely prior-dominated. Within these broad prior ranges, we find that uncertainties in the FRB redshift distribution have a relatively minor effect on the inferred cosmological and baryonic parameters. This approach effectively assumes that the redshift distribution can be approximately calibrated from localised FRBs, with residual uncertainties propagated through $\alpha$ and $\delta z$.
  
\begin{figure}
    \centering
    \includegraphics[width=\linewidth]{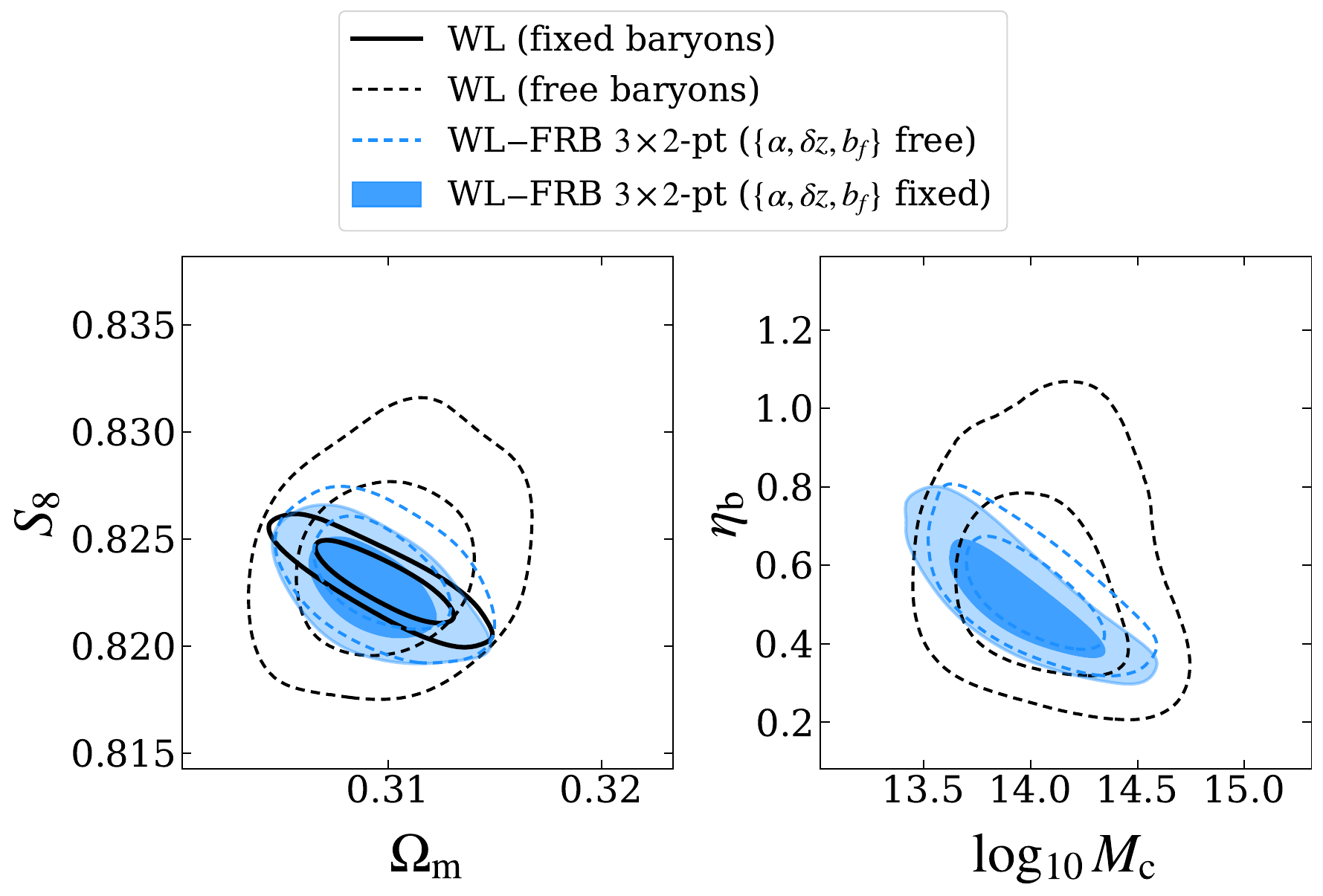}
    \caption{The marginalised posteriors on cosmological and baryonic parameters for LSST-like weak lensing data only for fixed baryonic parameters (solid black), LSST-like weak lensing data only with marginalisation over baryonic parameters (dashed black), and joint analyses of LSST-like weak lensing data with FRB DM correlations in a 3$\times$2-point analysis. The filled blue contours show the ideal case in which FRB systematics are well known, whereas the dashed blue contours represent a more realistic case in which we marginalise over FRB systematics.
    }
    \label{fig:combined_systematics}
  \end{figure}

  In Fig. \ref{fig:combined_systematics}, we provide an updated version of Fig. \ref{fig:ext_calibrator}, now including the marginalised posteriors obtained from the more complete treatment of FRB systematics. The solid blue contours correspond to the optimistic case in which $b_{\rm f}$, $\alpha$, and $\delta z$ are fixed to their fiducial values, while the dashed blue contours show the constraints obtained after marginalising over $b_{\rm f}$, $\alpha$, and $\delta z$. We find that adopting a more realistic modelling approach leads to a modest reduction in constraining power relative to the optimistic case, with the degradation factor for $S_8$ increasing from 1.2 to 1.3. Despite this, the combined 3$\times$2-point analysis of FRBs and cosmic shear continues to provide a significant improvement in cosmological constraints relative to cosmic shear alone, as highlighted by the reduction of degradation factor from 2.2 to 1.3, corresponding to an improvement of $\sim 80\%$. Moreover, we find that the majority of the loss in constraining power arises from marginalisation over FRB redshift uncertainties rather than from FRB source clustering effects. In particular, the degradation factor remains as 1.3 when the FRB bias $b_{\rm f}$ is held fixed to its fiducial value. This indicates that source clustering contributions are negligible, in agreement with Fig. \ref{fig:cls_source_clustering}. 

  Furthermore, the right-hand panel of Fig. \ref{fig:combined_systematics} shows that the FRB DM is primarily sensitive to a specific combination of $\log_{10} M_{\rm c}$ and $\eta_{\rm b}$, as discussed in the previous section. While the inclusion of FRBs reduces the allowed parameter volume in the $\log_{10} M_{\rm c}$--$\eta_{\rm b}$ plane, the constraint on $\log_{10} M_{\rm c}$ itself remains largely driven by WL rather than by the FRB data. As a result, we find that a joint 3$\times$2-point analysis of WL and FRBs yields substantially tighter constraints on baryonic feedback parameters than WL alone. In the optimistic scenario in which FRB systematics are calibrated, the constraints on $\log_{10} M_{\rm c}$ and $\eta_{\rm b}$ improve by factors of 1.1 and 1.7, respectively, relative to WL-only constraints. Hence, the inclusion of WL is essential for breaking the $\log_{10} M_{\rm c}$--$\eta_{\rm b}$ degeneracy.

  We also note that FRB-related systematics primarily act to broaden the constraints in the $\log_{10} M_{\rm c}$--$\eta_{\rm b}$ plane. Quantitatively, marginalising over the FRB systematic parameters degrades the constraints on $\log_{10} M_{\rm c}$ and $\eta_{\rm b}$ by factors of approximately 0.9 and 1.1, respectively, compared to the idealised case in which these parameters are fixed to their fiducial values.

  \subsubsection{Milky Way Foreground} \label{sssec:results.systematics.mw}
   As discussed in Section \ref{sssec:model.frb.survey}, the Milky Way contribution to the total DM acts as a foreground with a known spatial structure and is uncorrelated with the LSS. Its impact is therefore expected to be most significant on large angular scales. Here, we model this contribution as a large-scale systematic in order to explicitly account for potential contamination. Specifically, we impose a rather large minimum multipole scale-cut of $\ell_{\rm min}^{\mathcal{DD}} = 100$ in the FRB auto-correlation and compare the resulting constraints to the fiducial analysis with $\ell_{\rm min}^{\mathcal{DD}} = 2$. We find that introducing this scale-cut in the FRB mock data has a relatively mild impact on the inferred posterior constraints, with the degradation factor on $S_8$ increasing from 1.2 to 1.5. Given the modest impact of this scale-cut, we adopt the fiducial analysis with $\ell_{\rm max}^{\mathcal{DD}} = 2$ for the remainder of this work.
  
  \subsubsection{Effect of WL Multipole Range} \label{sssec:results.systematics.wl}
   Furthermore, we investigate the sensitivity of the cosmological constraints to the choice of maximum multipole in the WL mock data. To this end, we extend the multipole range of the LSST-like WL data from $\ell_{\rm max} = 2000$ to $\ell_{\rm max} = 5000$. We find that increasing the maximum multipole up to $\ell_{\rm max} = 5000$ leads to tighter constraints relative to the fiducial case. In particular, the degradation factor on $S_8$ for the WL--FRB 3$\times$2-point analysis decreases from 1.2 for $\ell_{\rm max} = 2000$ to 1.1 for $\ell_{\rm max} = 5000$. This improvement reflects the additional small-scale information incorporated into the analysis. However, the robustness of these results depends strongly on the assumed feedback model; future work is therefore required to assess the validity of the modelling on very small scales (e.g. $\ell_{\rm max} = 5000$).

   \begin{figure}
     \centering
     \includegraphics[width=\linewidth]{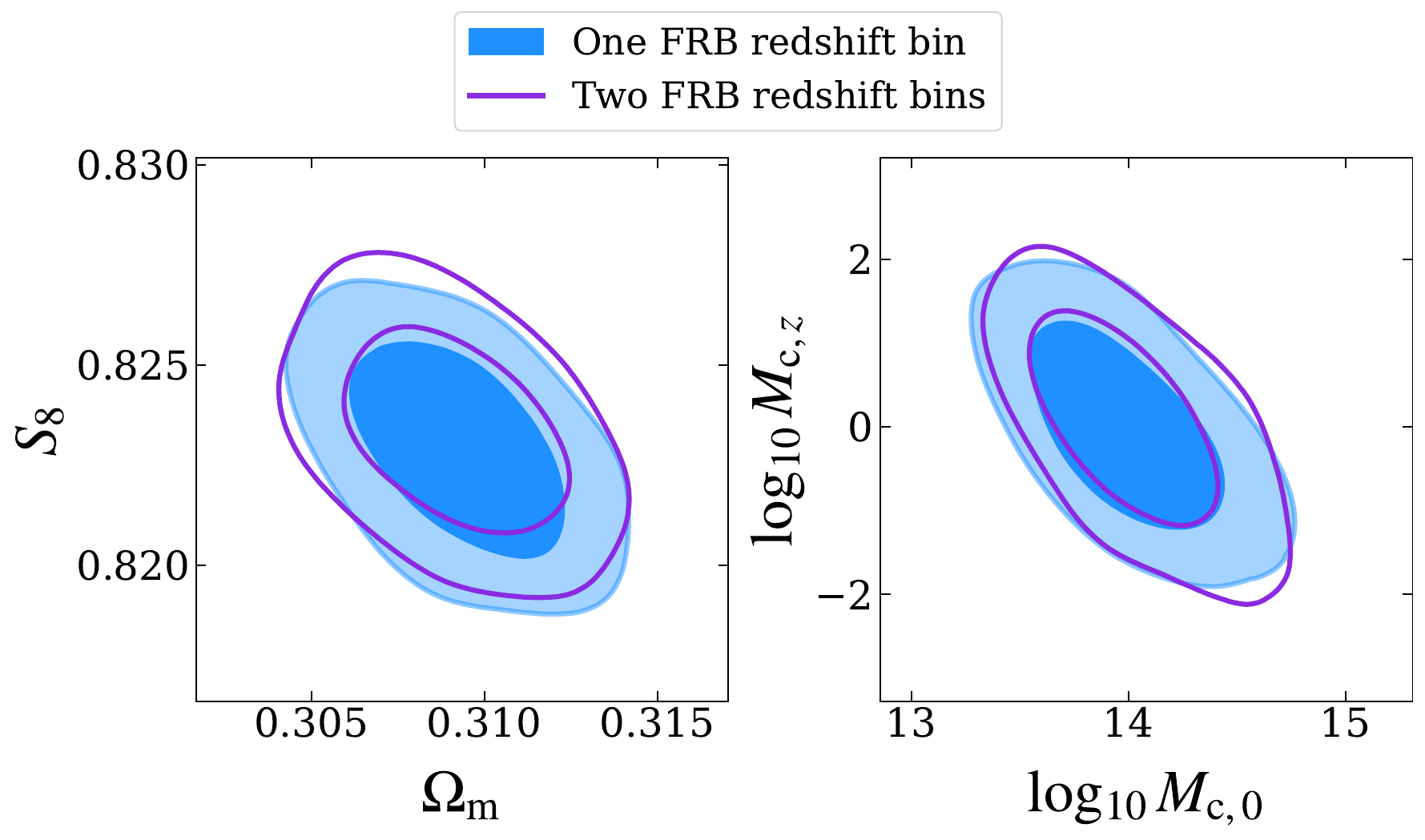}
     \caption{The marginalised posteriors on cosmological parameters and on the redshift dependence of $\log_{10} M_{\rm c}$ for LSST-like WL data combined with FRBs in a 3$\times$2-point analyses for two cases: (i) treating the full FRB redshift distribution as a single bin (blue), and (ii) dividing the FRB redshift distribution into two FRB tomographic bins (purple).
     }
     \label{fig:frb_tomography}
   \end{figure}
   
  \subsection{FRB tomography}
   Thus far, we have assumed a single FRB bin, motivated by the practical difficulty of obtaining redshifts for all sources. Nevertheless, it is informative to consider the potential gains from dividing the sample in redshift. Such a division could be achieved if a sufficient number of FRBs are localised, or alternatively by binning in DM. This approach may allow for the study of the potential redshift evolution of baryonic parameters. In this section, we consider redshift-dependent evolution of $\log_{10} M_{\rm c}$, since this parameter drives the amplitude of the power spectrum suppression.

   We divide the FRB sample into two redshift bins, $n_{\rm FRB}(z \leq z_{\rm max})$ and $n_{\rm FRB}(z > z_{\rm max})$, where $z_{\rm max}$ is the redshift at which the distribution attains its maximum value. We then compute the auto-correlation of each bin and the cross-correlation between the two bins, as well as all corresponding correlations with WL. These observables are appended to the data vector, and we repeat the full 3$\times$2-point analysis described above. 

   To model a potential redshift dependence of $\log_{10} M_{\rm c}$, we adopt the parametrisation
   \begin{equation} \label{eq:lMc_z_dep}
       \log_{10} M_{\rm c} = \log_{10} M_{\rm c, 0} + (1-a) \log_{10} M_{\mathrm{c}, a},
   \end{equation}
   where $\log_{10} M_{\rm c, 0}$ and $\log_{10} M_{\mathrm{c}, a}$ are treated as free parameters.

   In Fig. \ref{fig:frb_tomography}, we present the marginalised posteriors on the cosmological parameters for two WL--FRB 3$\times$2-point configurations, after introducing redshift dependence in $\log_{10} M_{\rm c}$ according to Eq. \eqref{eq:lMc_z_dep}. We first consider a configuration in which the full FRB redshift distribution is treated as a single bin, with the corresponding constraints shown in blue. We then perform FRB tomography by splitting the FRB sample into two redshift bins, with the resulting constraints shown in purple.

   In the single-bin case, introducing redshift dependence in $\log_{10} M_{\rm c}$ has only a negligible impact on the parameter constraints, with the degradation factor on $S_8$ increasing modestly from 1.2 to 1.4 relative to the fiducial case without redshift dependence in $\log_{10} M_{\rm c}$. The same result is observed when the sample is divided into two FRB redshift bins, indicating that FRB tomography does not improve the constraints on $S_8$ for the setup considered here. However, the two-bin approach offers a $\sim 3\%$ improvement in the constraints on $\log_{10} M_{\rm c,0}$. Overall, we find that FRB redshift tomography does not significantly affect our results, therefore, we adopt the one-bin configuration for the remainder of our analysis.
  
 \subsection{Constraints from future FRB surveys}
\begin{figure}
    \centering
    \includegraphics[width=\linewidth]{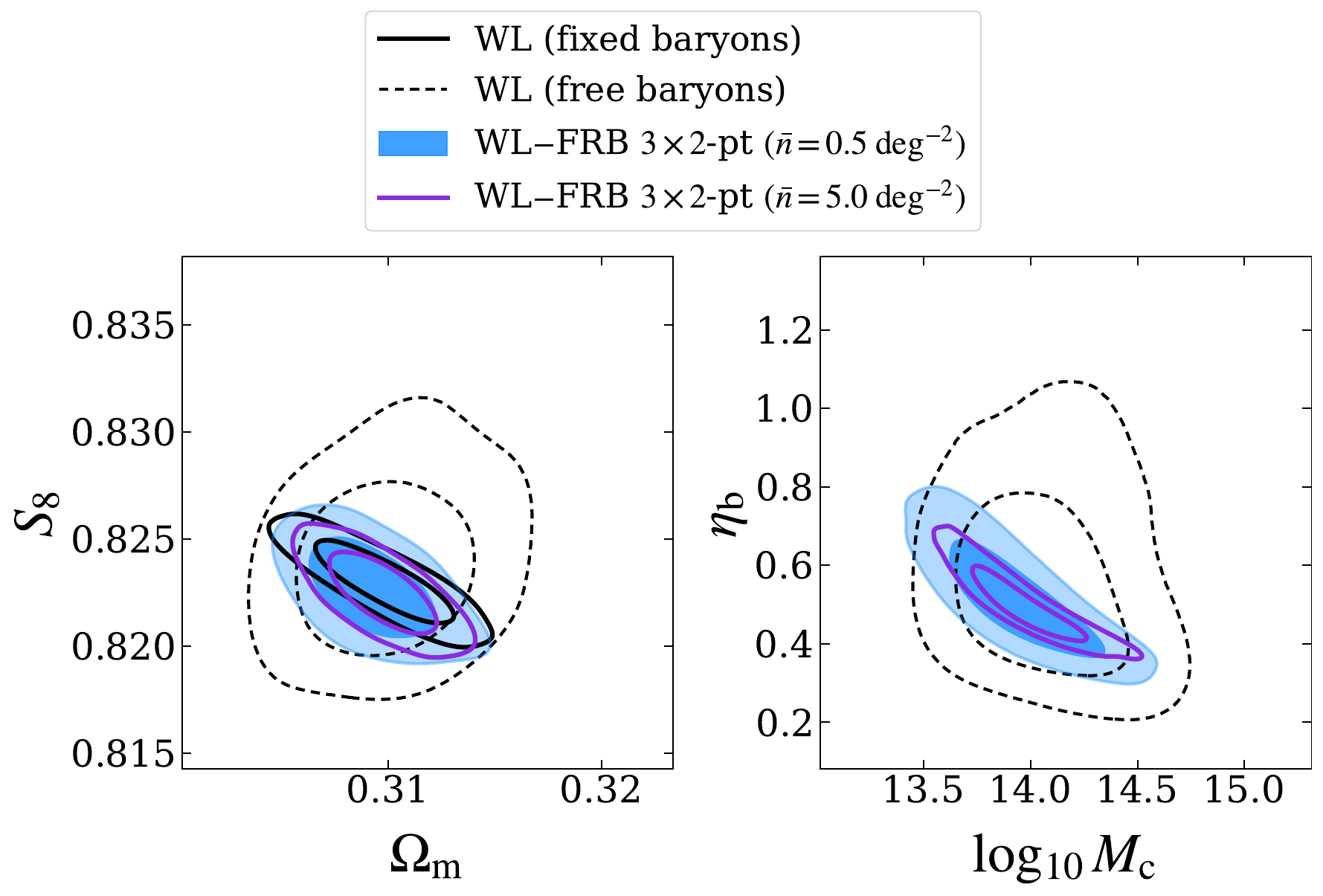}
    \caption{The marginalised posteriors on cosmological and baryonic parameters for LSST-like WL data only for fixed baryonic parameters (solid black), LSST-like WL data only with marginalisation over baryonic parameters (dashed black), and a joint 3$\times$2 analysis of LSST-like WL data with FRB DM correlations. The blue contours correspond to a near-term FRB sample with $\bar{n} = 0.5 \, \mathrm{deg}^{-2}$, while the purple contours correspond to a long-term FRB sample with $\bar{n} = 5.0 \, \mathrm{deg}^{-2}$.
    }
    \label{fig:future_survey}
  \end{figure}

  We now consider a more ambitious FRB sample, representative of what might be achieved by future surveys such as SKA \citep{Lazio2009SKA}, DSA-2000 \citep{Hallinan2019dsa}, and CHORD \citep{Vanderlinde2020CHORD}. To investigate this scenario, we increase the number density of FRBs by an order of magnitude, from $\bar{n} = 0.5 \, \mathrm{deg}^{-2}$ to $\bar{n} = 5.0 \, \mathrm{deg}^{-2}$. We then repeat the 3$\times$2-point analysis with WL, following the procedure outlined above, assuming the ideal case in which both the redshift distribution and source clustering effects are known. The resulting constraints are presented in Fig. \ref{fig:future_survey}. We find that the futuristic FRB sample yields a significant improvement in cosmological constraints, decreasing the degradation factor on $S_8$ from 1.2 to 1.0. Moreover, the constraint on $\log_{10} M_{\rm c}$ improves by a factor of 1.2, relative to the fiducial FRB sample with $\bar{n} = 0.5 \, \mathrm{deg}^{-2}$. These results highlight the value of longer-term FRB samples for enhancing the constraining power of joint WL--FRB analyses. 

 \subsection{6$\times$2-point analysis of FRB DMs, WL, and GC}
  In this section, we extend the 3$\times$2-point framework to include galaxy clustering in a unified 6$\times$2-point analysis. Specifically, we consider two galaxy populations, LRGs and ELGs. These galaxy samples probe different halo masses, thereby allowing us constrain the halo-mass dependence of the bound gas fraction, $f_{\rm bgas}(M)$. In turn, this could allow for more stringent constraints to be placed on the bound gas parameter $\log_{10} M_{\rm c}$. The LRG and ELG populations are modelled using the HOD and SFHOD models, respectively, as described in Section \ref{ssec:model.galaxies}.

\begin{figure}
    \centering
    \includegraphics[width=\linewidth]{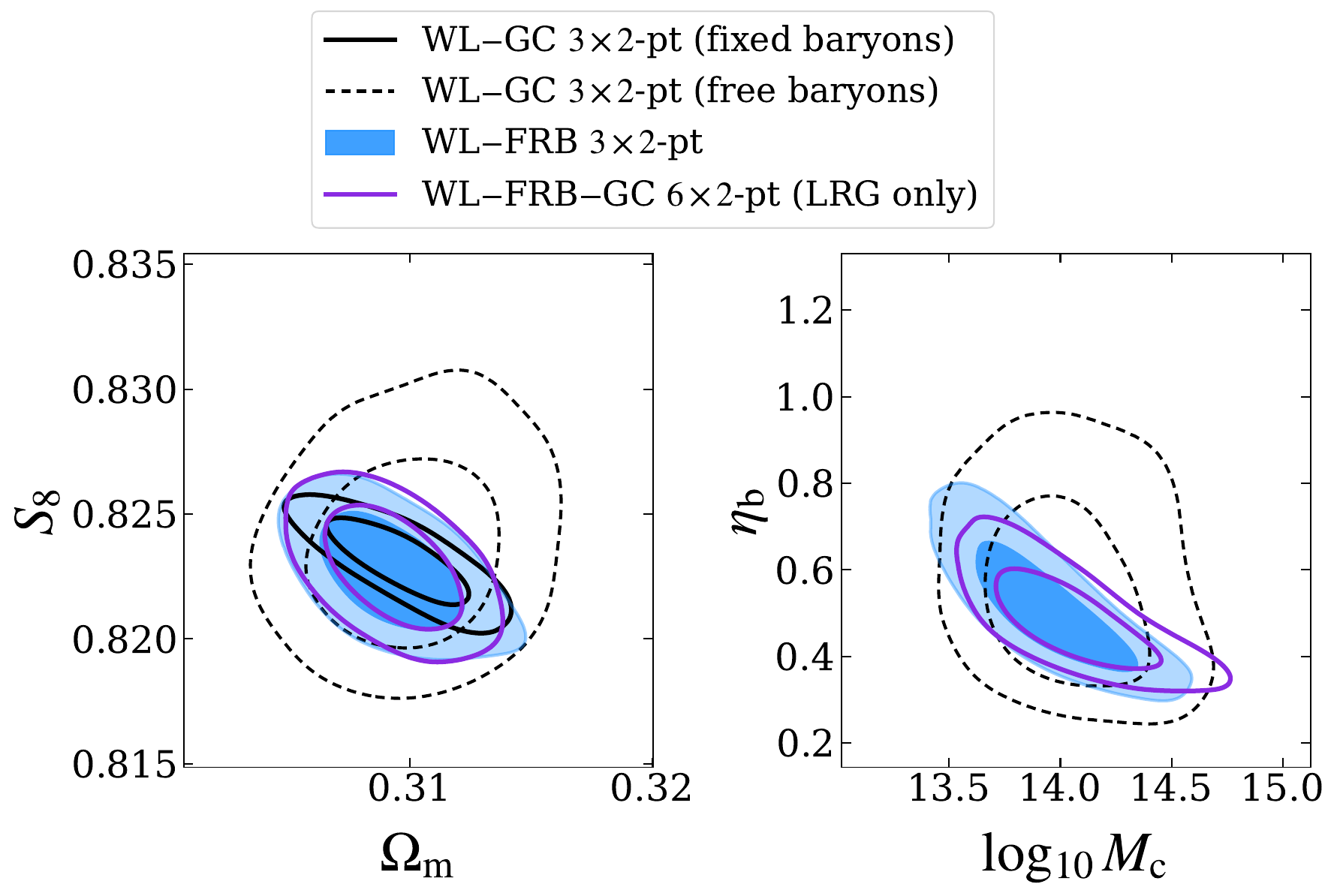}
    \caption{The marginalised posteriors on cosmological and baryonic parameters for different combinations of tracers. The solid black contours show the constraints obtained from the combined WL--GC 3$\times$2-point analysis under the assumption of perfect knowledge of baryonic effects, while the dashed black contours show those obtained from the combined WL--GC 3$\times$2-point analysis with marginalisation over baryonic effects. The blue contours show the constraints obtained from a 3$\times$2-point analysis of WL and FRBs, while the purple contours show those obtained from a unified 6$\times$2-point analysis of WL, FRBs, and GC. The latter involves a single galaxy tracer, assuming a photometric LRG sample.
    }
    \label{fig:6x2pt_lrg_only}
  \end{figure}

  We first consider a single galaxy tracer, assuming a photometric LRG-like sample. We marginalise over the HOD parameters $\log_{10} M_{\rm min}$ and $\log_{10} M_1$. As shown in Fig. \ref{fig:6x2pt_lrg_only}, incorporating this sample into a 6$\times$2-point analysis with WL and FRB DMs yields only a negligible improvement relative to the WL--FRB 3$\times$2-point analysis. Quantitatively, the degradation factor on $S_8$ decreases from 2.4 for the WL--GC 3$\times$2-point analysis to 1.4 for the WL--FRB 3$\times$2-point analysis, while the degradation factor remains at 1.4 for the full WL--FRB--GC 6$\times$2-point case. Similarly, there is a negligible change in the error on $\log_{10} M_{\rm c}$. This outcome is consistent with our expectations, as a single galaxy sample does not sufficiently probe the mass dependence of $f_{\rm bgas}(M)$, and therefore cannot meaningfully improve the constraints on the bound gas parameter $\log_{10} M_{\rm c}$. 

\begin{figure}
    \centering
    \includegraphics[width=\linewidth]{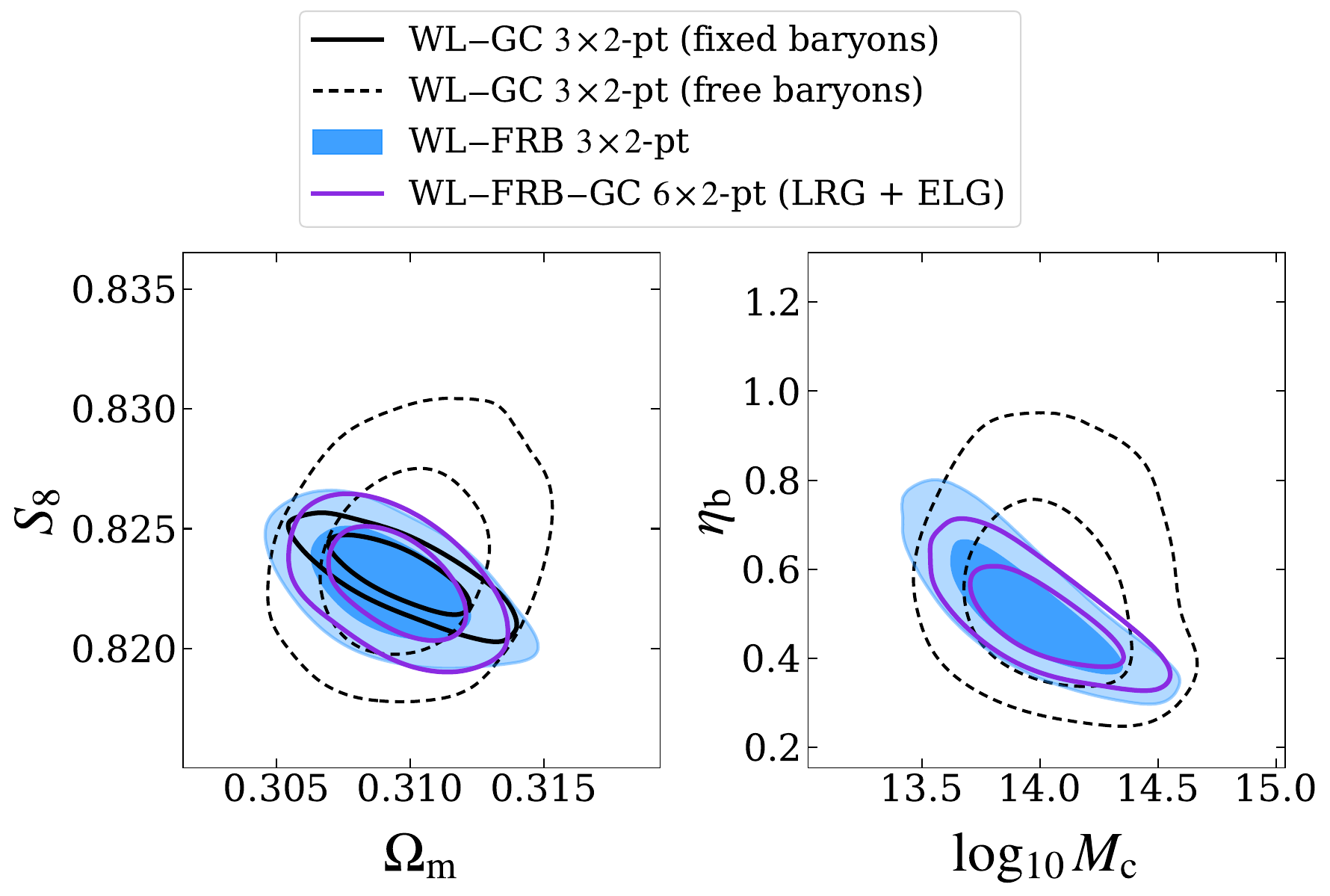}
    \caption{As Fig \ref{fig:6x2pt_lrg_only}, but now including both LRGs and ELGs, probing complementary halo mass ranges.
    }
    \label{fig:6x2pt_lrg_elg}
  \end{figure}

  We then incorporate an ELG sample as an additional galaxy tracer. For each galaxy sample, we marginalise over the HOD parameters $\log_{10} M_{\rm min}$ and $\log_{10} M_1$. As illustrated in Fig. \ref{fig:6x2pt_lrg_elg}, combining LRGs and ELGs in the full 6$\times$2-point analysis with WL and FRB DMs reduces the error on $\log_{10} M_{\rm c}$ by a factor of 1.1 relative to the WL--FRB 3$\times$2-point case. Interestingly, this improvement in constraining power on $\log_{10} M_{\rm c}$ does not propagate to tighter constraints on $S_8$. While the degradation factor on $S_8$ is reduced from 2.4 for the WL--GC 3$\times$2-point analysis to 1.4 for the WL--FRB 3$\times$2-point case, it remains at 1.4 when extending to the full WL--FRB--GC 6$\times$2-point analysis. Moreover, comparing the purple contours in the $S_8$--$\Omega_{\rm m}$ plane in Figs. \ref{fig:6x2pt_lrg_only} and \ref{fig:6x2pt_lrg_elg} shows that including ELGs as an additional galaxy sample improves the constraints on $\Omega_{\rm m}$ by a factor of 1.1. This suggests that the additional galaxy information primarily enhances sensitivity to $\Omega_{\rm m}$ and baryonic parameters, with limited impact on the cosmological degeneracies governing $S_8$.

\section{Conclusions} \label{sec:conclusions}
 In this work, we explored the potential of FRBs as tracers of large-scale structure and as a complementary probe to weak lensing and galaxy clustering in constraining both cosmological parameters and baryonic feedback. The angular power spectrum of the FRB DM encodes the electron auto-power spectrum, with a redshift sensitivity that peaks in the foreground of FRB sources. As a result, FRB DM perturbations are particularly sensitive to baryonic physics and cosmology through their dependence on the electron power spectrum, the mean DM-redshift relation, and their associated radial weighting function \citep{Sharma2025probing}.

 We investigated the cross-correlation between FRB DM perturbations and WL through their angular cross-power spectrum, which probes the electron-matter cross-power spectrum. By incorporating FRBs into a 3$\times$2-point analysis alongside LSST-like WL data, we showed that FRB--WL correlations provide additional information beyond WL alone, helping to break degeneracies between cosmological and baryonic parameters. Quantitatively, we found a substantial reduction in the degradation factor on the $S_8$ uncertainty, from 2.2 for WL alone to 1.2 for the WL--FRB 3$\times$2-point analysis. Crucially, FRB data is sensitive to a somewhat degenerate combination of the bound gas abundance (characterised by $\log_{10}M_{\rm c}$), and the spatial extent of the ejected gas (characterised by $\eta_{\rm b}$). This degeneracy is then partially broken through the combination with WL data, which exhibits stronger sensitivity to $\log_{10}M_{\rm c}$.

 Moreover, the inclusion of FRB DMs yields a significant gain in sensitivity to the Hubble parameter $h$. In contrast to WL, which depends only weakly on $h$, FRB DMs probe the distance-redshift relation directly, leading to a characteristic response of both the FRB auto-correlation and the WL--FRB cross-correlation that is distinct from the effects of baryonic feedback. This enables the joint analysis to substantially tighten constraints on $h$, even when baryonic parameters are allowed to vary. Since extending the cosmological parameter space does not otherwise alter our main conclusions, we therefore adopt a simplified cosmological model in which $\Omega_{\rm m}$ and $\sigma_8$ are varied for the remainder of the analysis, while highlighting the improved constraint on $h$ as a key additional gain enabled by FRB data.

 We then examined the key systematics relevant to FRB surveys, including uncertainties in the FRB redshift distribution and source clustering of FRBs with their host galaxies. These effects were modelled through additional nuisance parameters describing photometric redshift uncertainties and the FRB bias. We found that such systematics lead only to a modest degradation in cosmological constraints, with the degradation factor on $S_8$ increasing from 1.2 to 1.3 when these uncertainties are included. Instrumental calibration effects associated with the DM selection function remain an important systematic for future work \citep[see e.g.][]{Rafiei2019characterising, Rafiei2021CHIME}. For instance, the instrument DM selection function may prevent the detection of FRBs with high DMs, which would in turn bias the constraints on baryonic feedback parameters \citep{Cheng2025exploring}.

 We further considered several extensions to our baseline analysis. For a more futuristic FRB survey with an order-of-magnitude increase in FRB number density, we found a corresponding improvement in cosmological constraints, with the degradation factor on $S_8$ further decreasing from 1.2 to 1.0. This indicates that a 3$\times$2-point analysis with such a future FRB survey could fully recover the loss of information in the $S_8$ constraint induced by uncertainties surrounding baryonic feedback. Moreover, we assessed the ability of FRB tomography to constrain a possible redshift dependence of the parameter $\log_{10} M_{\rm c}$ by dividing the FRB redshift distribution into two tomographic bins. However, this tomographic approach did not yield a significant improvement in the constraining power on $S_8$ relative to the single-bin case. Comparing FRBs to other probes of the gas distribution, we found that FRBs alone exhibit the aforementioned degeneracy between the baryonic parameters $\log_{10} M_{\rm c}$ and $\eta_{\rm b}$. In contrast, mock X-ray and kSZ measurements based on the eROSITA survey and a CMB-S4-like experiment can place stringent constraints on $\log_{10} M_{\rm c}$ and $\eta_{\rm b}$, respectively, highlighting the complementary nature of these observables. 

 Finally, we extended our analysis to include galaxy clustering, performing a full 6$\times$2-point analysis combining FRBs, WL, and galaxy clustering samples. By incorporating both luminous red galaxies and emission line galaxies, we were able to probe the bound gas fraction across different halo mass scales and obtain tighter constraints on $\Omega_{\rm m}$ and the bound gas parameter $\log_{10} M_{\rm c}$. Although galaxy auto-correlations are highly sensitive to cosmological and halo occupation parameters, they are largely insensitive to baryonic feedback. In contrast, FRBs directly trace the spatial distribution of ionised gas and therefore provide direct sensitivity to baryonic parameters. Despite this complementarity, we found that the full 6$\times$2-point analysis did not yield any significant improvement in the constraints on $S_8$ compared to the 3$\times$2-point analysis combining FRBs and WL.

 A potential avenue for future work is to extend our framework by jointly analysing the FRB DM--redshift relation together with the FRB angular power spectra. While our analysis has focused on DM fluctuations, the mean DM--$z$ relation encodes complementary information on the integrated electron density and its redshift evolution, and can further help to break degeneracies between cosmological parameters and baryonic feedback \citep{Reischke2025first, Sharma2025hydrodynamical, Medlock2025constraining}. A consistent treatment that combines DM perturbations and the DM--$z$ relation, including the relevant systematics, would therefore strengthen the cosmological and astrophysical interpretation of FRB observations. In addition, pushing the analysis to smaller angular scales would allow more stringent tests of baryonic feedback models, where differences between feedback prescriptions are expected to be more pronounced. Such an extension will require improved modelling of non-linear gas astrophysics and potential limitations of the halo-based description adopted here. Together, these developments would enable FRBs to more fully realise their potential as probes of baryonic physics and structure formation across a wide range of scales.

 Overall, our results highlight the complementary role of FRBs in multi-tracer analyses. As FRB samples continue to grow and are combined with next-generation WL and galaxy surveys, such joint analyses could play a crucial role in disentangling baryonic effects and on improving the precision of specific cosmological parameters, enabling more robust constraints to be obtained in the era of precision cosmology.

\section*{Acknowledgements}
  We thank Steffen Hagstotz and Andrina Nicola for useful discussions. AW is supported by a Science and Technology Facilities Council studentship. DA acknowledges support from the Beecroft Trust. We made extensive use of computational resources at the University of Oxford Department of Physics, funded by the John Fell Oxford University Press Research Fund.

\section*{Data Availability}
 The data underlying this article will be shared on reasonable request to the corresponding authors.

\bibliographystyle{mnras}
\bibliography{references}

\appendix
\section{Intrinsic host DM fluctuations}\label{app:host_more}
  We present here a simplified model illustrating the presence of the contribution from intrinsic host DM fluctuations, ${\cal D}_{\rm host}$, described in Section \ref{sssec:model.frb.sc}.

  Consider a simplified halo model for FRBs in which haloes may host 0 or 1 FRBs at their centre, with $N_{\rm f}(M)$ the expected number of FRBs in a halo of mass $M$. Given a model for the gas content in haloes, we can then calculate the DM of the central FRB due to the host halo, ${\rm DM}_{\rm host}(M,z)$. For simplicity, let us consider this to be the only contribution to the observed DM, ignoring that of the intervening IGM. The host ``DM density'' (i.e. cumulative host DM per unit volume) for FRBs at redshift $z$ and position ${\bf x}$ can then be calculated as
  \begin{equation}
    \rho_{\rm DM}({\bf x})=\int \mathrm{d}M\,n(M)\,{\rm DM}_{\rm host}(M)\,N_{\rm f}(M)\,[1+\delta_{\rm h}({\bf x}|M)],
  \end{equation}
  where $n(M)$ is the halo mass function, and $\delta_{\rm h}({\bf x}|M)$ is the overdensity of haloes of mass $M$. For ease of notation we have omitted the redshift dependence of all quantities, and we have ignored the factor of $1/(1+z)$ relating the ${\rm DM}$ in the rest-frame of the FRB to the observer. Likewise, the number density of FRBs is
  \begin{equation}
    n_{\rm f}({\bf x})=\int \mathrm{d}M\,n(M)\,N_{\rm f}(M)\,[1+\delta_{\rm h}({\bf x}|M)].
  \end{equation}
  The angular DM and FRB number densities (i.e. cumulative host DM and number of FRBs per unit solid angle) are then given by
  \begin{equation}\nonumber
    \Sigma_{\rm DM}(\nv)=\int \mathrm{d}\chi\,\chi^2\rho_{\rm DM}(\chi\nv),\hspace{6pt}
    n_{\Omega,f}(\nv)=\int \mathrm{d}\chi\,\chi^2\,n_{\rm f}(\chi\nv).
  \end{equation}
  The observed DM map is then the ratio of these two quantities ${\rm DM}_\Omega(\nv)=\Sigma_{\rm DM}(\nv)/n_{\Omega,f}(\nv)$. Expanding this to linear order in the halo overdensity, we obtain
  \begin{align}\nonumber
    &{\rm DM}_\Omega(\nv)=\overline{\rm DM}_\Omega + \\\nonumber
    &\hspace{5pt}\int \frac{\mathrm{d}\chi\,\chi^2}{\bar{n}_{\Omega,f}}\int \mathrm{d}M\,n(M)N_{\rm f}(M)\left[{\rm DM}_{\rm host}(M)-\overline{\rm DM}_\Omega\right]\delta_{\rm h}(\chi\nv|M),
  \end{align}
  where the mean projected ${\rm DM}$ and FRB angular number density are
  \begin{align}\nonumber
    \bar{n}_{\Omega,f}&\equiv\int \mathrm{d}\chi\,\chi^2\int \mathrm{d}M\,n(M)\,N_{\rm f}(M)\\\nonumber
    &\equiv\int \mathrm{d}\chi\,\chi^2\,\bar{n}_{\rm f}(\chi),\\\nonumber
    \overline{\rm DM}_\Omega&\equiv \frac{1}{\bar{n}_{\Omega,f}}\int \mathrm{d}\chi\,\chi^2\int \mathrm{d}M\,n(M)\,N_{\rm f}(M)\,{\rm DM}_{\rm host}(M)\\\nonumber
    &\equiv\frac{1}{\bar{n}_{\Omega,f}}\int \mathrm{d}\chi\,\bar{n}_{\rm f}(\chi)\,\overline{\rm DM}(\chi),
  \end{align}
  which also defines the mean FRB number density, $\bar{n}_{\rm f}(\chi)$, and the mean host ${\rm DM}$ at comoving distance $\chi$, $\overline{\rm DM}(\chi)$.
  
  Assuming haloes to be linearly biased tracers, $\delta_{\rm h}({\bf x}|M)\simeq b_h(M)\,\delta_{\rm m}({\bf x})$, we can write the angular fluctuation in host DM ${\cal D}(\nv)\equiv {\rm DM}_\Omega(\nv)-\overline{\rm DM}_\Omega$ as
  \begin{equation}\label{eq:dhost_gen}
    {\cal D}(\nv)=\int \frac{\mathrm{d}\chi\,\chi^2}{\bar{n}_{\Omega,f}}\,\bar{n}_{\rm f}(\chi)\left[b_{\rm f,{\rm DM}}\,\overline{\rm DM}(\chi)-b_{\rm f}\overline{\rm DM}_\Omega\right]\delta_{\rm m}(\chi\nv),
  \end{equation}
  where we have defined the FRB bias, $b_{\rm f}$, and the DM-weighted FRB bias, $b_{\rm f,{\rm DM}}$, as
  \begin{align}
    &b_{\rm f}\equiv\frac{\int \mathrm{d}M\,n(M)\,N_{\rm f}(M)\,b_h(M)}{\int \mathrm{d}M\,n(M)\,N_{\rm f}(M)},\\
    &b_{\rm f,{\rm DM}}\equiv\frac{\int \mathrm{d}M\,n(M)\,N_{\rm f}(M)\,{\rm DM}(M)\,b_h(M)}{\int \mathrm{d}M\,n(M)\,N_{\rm f}(M)\,{\rm DM}(M)}.
  \end{align}

  We can see that Eq. \eqref{eq:dhost_gen} is equal to the last term in Eq. \eqref{eq:DM_with_host} if $b_{\rm f}=b_{\rm f,{\rm DM}}$, which is not in general the case, due to the expected mass dependence of the host DM. Thus, the intrinsic fluctuation in the host DM contribution ${\cal D}_{\rm host}$ in Eq. \eqref{eq:DM_with_host} is given, in this simplified model, by:
  \begin{equation}
    {\cal D}_{\rm host}(\nv,\chi)=\left[b_{\rm f,{\rm DM}}-b_{\rm f}\right]\,\overline{\rm DM}(\chi)\,\delta_{\rm m}(\chi\nv).
  \end{equation}
  

\bsp	
\label{lastpage}
\end{document}